\documentclass[%
 reprint,
superscriptaddress,
amsmath,amssymb,
aps,
prl,
]{revtex4-2}

\usepackage{graphicx}% Include figure files
\usepackage{dcolumn}% Align table columns on decimal point
\usepackage{bm}% bold math
\usepackage{easyReview}
\usepackage{physics}
\usepackage{dsfont}
\usepackage{leftindex}
\usepackage{times}
\usepackage{orcidlink}
\usepackage{hyperref}
\usepackage{subfigure}
\usepackage{xcolor}
\hypersetup{
    colorlinks=true,
    citecolor=blue,
    linkcolor=black,
    filecolor=magenta,      
    urlcolor=blue,
    pdfpagemode=FullScreen,
    }

\begin{document}
\title{Gravitational Wave-Induced Superradiance in Ordered Atomic Arrays}% 
\author{Navdeep Arya\orcidlink{https://orcid.org/0000-0003-0730-4835}}
\affiliation{Department of Physics, Stockholm University, SE-106 91 Stockholm, Sweden}
\author{Magdalena Zych\orcidlink{0000-0002-8356-7613}}
\affiliation{Department of Physics, Stockholm University, SE-106 91 Stockholm, Sweden}
\affiliation{ARC Centre for Engineered Quantum Systems, School of Mathematics and Physics, The University of Queensland, St. Lucia, Queensland, 4072, Australia}
	\begin{abstract}
		The effects of spacetime geometry on quantum systems are typically very small. Here, we demonstrate a coherent many-body mechanism that can enhance these effects. We show that, in an ordered array, a gravitational wave induces long-range all-to-all dissipative coupling among atoms within half the gravitational wavelength. This coupling is mediated by the electromagnetic vacuum and leads to cooperative photon emission that we term \textit{gravitational wave-induced photon superradiance}—delayed and intense emission of photons at frequencies shifted from the atomic transition by the gravitational wave frequency. The phenomenon arises in a regime distinct from flat-spacetime superradiance, allowing gravitational effects to dominate the collective photon emission from atoms. It persists despite common experimental challenges in atom arrays such as position disorder and partial filling. We thus identify a new class of effects arising from the interplay of general relativity and collective quantum optics that individual atoms do not exhibit, and demonstrate that engineered quantum many-body systems provide a new window into the interface of general relativity and quantum mechanics. 
	\end{abstract}
  \maketitle  
  %%%%%%%%%%%%%%%%%%%%%%%%%%%%%%%%%%%%%%%%%%%%%%%%%%%%%%%%%%%%%%%%%%%%
\paragraph*{Introduction---}
Gravitational effects beyond the nonrelativistic effect of free fall are challenging to measure because of the weak nature of gravity compared to other fundamental interactions. Despite growing efforts to understand quantum aspects of gravity, even the coupling between general relativistic gravity and quantum matter has not yet been probed experimentally. Therefore, witnessing effects that would require both quantum physics and general relativistic gravity for their description is an area of growing interest. This interest is further underscored by the fact that ever-growing precision in quantum measurement and sensing techniques is expected to reach the sensitivity where general relativistic effects will have to be included in their design and description to reach their target performance~\cite{Bothwell2022,Zheng2023}. As a result, the development of both theoretical tools and experimental proposals that aim to bridge quantum physics and general relativistic gravity is gaining momentum~\cite{Zych2011,Hohensee2012,Pikovski2015,Asenbaum2017,Loriani2019,Roura2020,Overstreet2022,Paczos2024a,Paczos2025a,AnjunChu2025,Borregaard2025,Covey2025,Paczos2026}.

Here, we identify a new coupling interface between general relativistic classical gravity and quantum matter: spacetime curvature introduced by a gravitational wave~(GW) can qualitatively reshape collective photon emission from an ordered array of atoms.
In flat spacetime, an array of atoms with interatomic spacing much less than their transition wavelength can emit photons in a delayed, short, and intense burst---a process known as superradiant emission~\cite{Dicke1954,Rehler1971,Eberly1972,Gross1982,Masson2020,Masson2022}.
In free space, however, the peak total intensity scales quadratically with the number of excited atoms only in the ideal Dicke limit, where the sample size is much smaller than the atomic transition wavelength~\cite{Dicke1954}. As collective emission sensitively depends on the spatial distribution of atoms and on the properties of the quantum electromagnetic (EM) field to which they couple~\cite{Deswal&Arya2025}, modifications of the field by spacetime curvature can alter the cooperative behavior of the array. 

We find that in the presence of a GW, the condition that ensures cooperative behavior of atoms changes qualitatively as the spacetime curvature affects the processes mediated by the EM field: while superradiant emission in flat spacetime is strong for interatomic spacings $(d)$ much smaller than the atomic transition wavelength $(\lambda_0)$, the GW-induced cooperative emission dominates for $d \approx \lambda_0$. This sharp separation of regimes provides a setting in which the total collective photon emission from the atoms can be exclusively driven by the GW and we therefore refer to this effect as \textit{GW-induced superradiance}.

\begin{figure}
    \centering
    \includegraphics[width=1\linewidth]{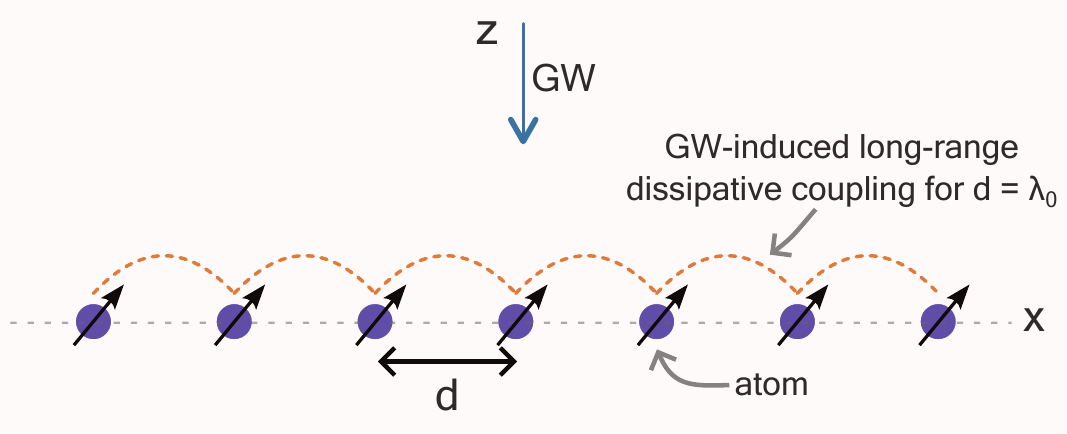}
    \caption{Schematic depiction of the envisioned scenario.}
    \label{fig:schematic}
\end{figure}

The effect reported here is not only genuinely quantum-mechanical and general-relativistic, but also intrinsically many-body. An initially uncorrelated array of fully excited atoms becomes sensitive to the GW only after quantum coherence spontaneously builds up among the atoms, seeded by vacuum fluctuations of the EM field.
This collective sensitivity stands in sharp contrast to individual atoms, whose total emission rates remain insensitive to gravitational waves at first order in the wave's amplitude~\cite{Qidong2020,Gray2021,Prokopec2023,Barman2023,Paczos2026}.
The underlying mechanism is a GW-induced long-range all-to-all dissipative coupling between the atoms, analogous to the coupling mediated by a cavity or waveguide~\cite{Goban2015,Solano2017}. It is this coupling that gives rise to GW-induced superradiance---emission of photons at frequencies $|\omega_0 \pm \omega|$, shifted from the atomic transition frequency $\omega_0$ by the GW frequency~$\omega$, with peak intensity scaling quadratically with the number of excited atoms. The emitted photon intensity beats at the GW frequency, with the beat profile encoding the phase of the wave.
We use the $\hbar = c = 1$ convention, however $c$ is restored on occasion for clarity.
%%%%%%%%%%%%%%%%%%%%%%%%%%%%%%%%%%%%%%%%%%%%%%%%%%%
%%%%%%%%%%%%%%%%%%%%%%%%%%%%%%%%%%%%%%%%%%%%%%%%%%%
\paragraph*{Dissipative coupling and photon emission---}
Consider $N$ identical two-level atoms, with transition frequency $\omega_0$, arranged in a one-dimensional (1D) array with interatomic distance $d$ along $x$-direction [see Fig.~\ref{fig:schematic}]. 
A plane linearized GW of frequency $\omega$, wavelength $\lambda_{\rm gw}$, polarization $+$, and amplitude $h_{+}$ is assumed to propagate transverse to the array. The mode structure of the quantum EM field, to which the atoms couple, is modified by the GW-background spacetime~\cite{SM}. 
%%%%%%%%%%%%%%%%%%%%%%%%%%%%%%%%%%%%%%%%%%%%%%%%%%%%%%%%%%%%%%%%%%%%%%%%%%%%
\begin{figure*}
	\centering
	\includegraphics[width=0.99\linewidth]{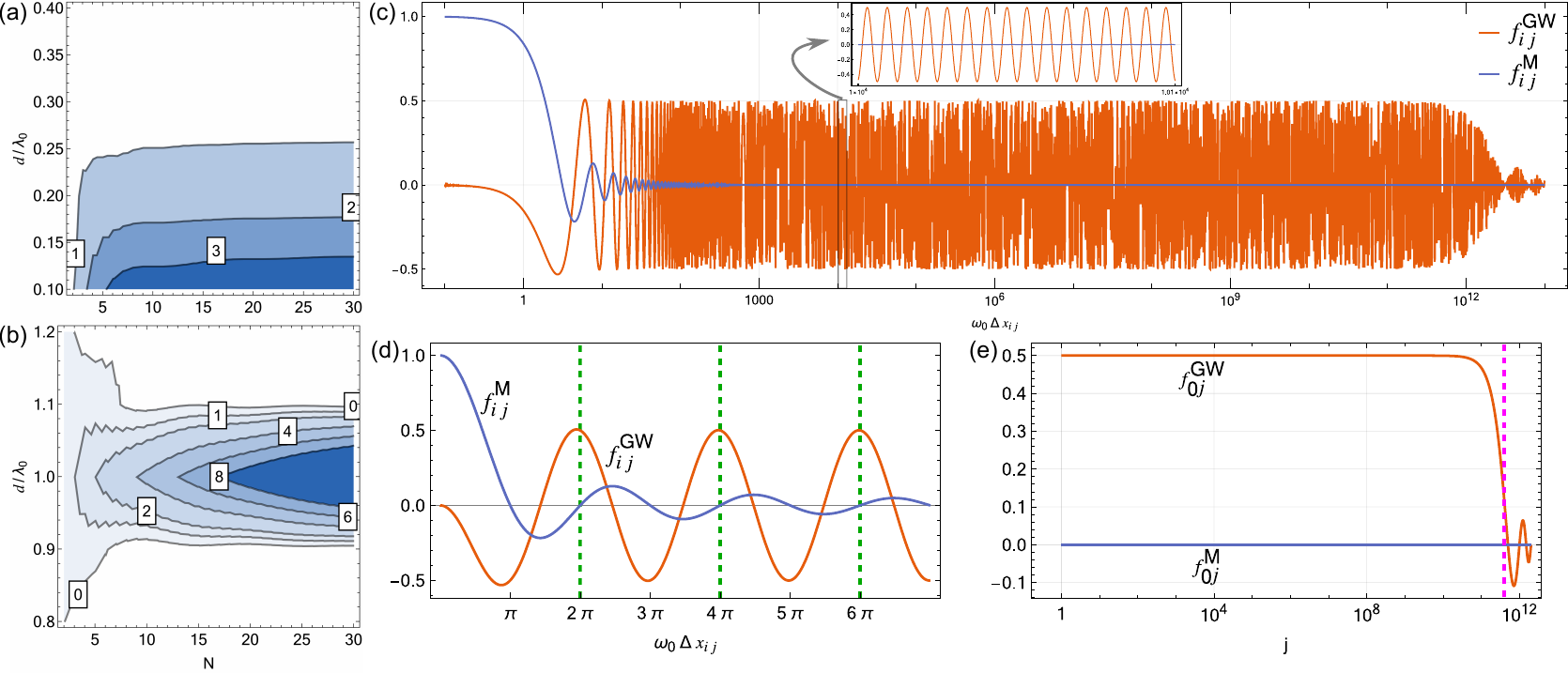}
	\caption{(a) and (b) depict the regimes of flat-spacetime and GW-induced superradiances. 
    \textbf{(a)} Contour plot showing constant values of $\mu_{\rm M}N$, the flat-spacetime cooperation among the atoms, as a function of the number of atoms $N$ and the ratio $d/\lambda_0$. For a 1D array of atoms to show a superradiant burst in flat spacetime requires $\mu_{\rm M} N > 1$. In flat spacetime a superradiant burst survives for $d \lesssim 0.25 \lambda_0$. \textbf{(b)} Contour plot showing constant values of the GW-induced cooperation among the atoms, $\eta \equiv |\mu_{\rm GW}(0)| N/h_{+}$. Large values of $\eta$ can be achieved for $d \approx \lambda_0$. In contrast, the flat-spacetime contribution to collective response is highly suppressed in this regime (not shown). \textbf{(c)} Comparison of functions $f_{ij}^{\rm M}$ and $f_{ij}^{\rm GW}$ that govern the collective dissipative response of atoms [see Eq.\,\eqref{Fij}]. For two atoms with $\omega_0 \Delta x_{ij} \gtrsim \pi$, $f_{ij}^{\rm M}$ is highly suppressed whereas the GW enables dissipative coupling between distant atoms. In the portion that appears smeared, $f_{ij}^{\rm GW} \sim \cos(\omega_0 \Delta x_{ij})$ as shown in the inset. The plot is for $\bar{\omega} = 10^{-12}.$ \textbf{(d)} A closer look at the two functions. The dashed green lines correspond to the atomic positions prescribed for enhancing the GW-induced dissipative coupling and suppressing the flat-spacetime dissipative coupling. \textbf{(e)} Plots of $f^{\rm GW}_{0j}$ and $f^{\rm M}_{0j}$ as a function of $j$, for $d/\lambda_0 =1$. For a given $\bar{\omega}$, the reference atom, here $i_0 = 0$,  is dissipatively coupled to atoms on other sites $j \leq j_{\rm max} \approx \lambda_{\rm gw}/2\lambda_0$. For a high enough value of $j = j_{\rm max}$, $f^{\rm GW}_{0j}$ diminishes and flips between positive and negative values beyond $j_{\rm max}$, and stays small. The first instance of this sign flip, illustrated by the dashed magenta vertical line, marks the maximum effective number of atoms, $\eta_{\rm max}$, showing the GW-induced cooperation. For $\bar{\omega} = 10^{-12}$, $\eta_{\rm max} \sim 10^{11}$.
    }
	\label{fig:gwmink-kernels}
\end{figure*}
%%%%%%%%%%%%%%%%%%%%%%%%%%%%%%%%%%%%%%%%%%%%%%%%%%%%%%%%%%%%%%%

%%%%%%%%%%%%%%%%%%%%%%%%%%%%%%%%%%%%%%%%%%%%%%%%%%%
Atoms coupled to a common EM field can influence each other's dynamics through field-mediated interactions. In particular, the modification of the spontaneous emission rates due to the presence of neighboring atoms depends on $\gamma_0 F_{ij}$ which is known as the dissipative coupling rate between atoms $i$ and $j$~\cite{Asenjo-Garcia2017}. Here, $\gamma_0$ is the spontaneous emission rate of a single atom in free space, and $F_{ij}$ is a function of the atomic transition frequency $\omega_0$, separation $\Delta x_{ij} = x_i - x_j$ between the $i$th and $j$th atoms, gravitational wave's amplitude $h_{+}$, and frequency $\omega$. To cleanly identify the features of interest, we first work in the scalar-light model before considering the full EM model. Both the models are discussed in the Supplemental Material~\cite{SM}. The $F_{ij}$ is given as~\cite{SM}:
\begin{equation}\label{Fij}
    F_{ij}(t) \equiv f^{\rm M}_{ij}(\omega_0 \Delta x_{ij}) + h_+ \cos(\omega t) f^{\rm GW}_{ij}\left(\omega^{\pm}_0 \Delta x_{ij} \right),
\end{equation}
where $\omega^{\pm}_0 \equiv \omega_0 \pm \omega$, $f_{ij}^{\rm M}(\omega_0 \Delta x_{ij}) \equiv {\rm sinc}(\omega_0 \Delta x_{ij})$ is the flat-spacetime (Minkowskian) dissipative coupling and the second term, with
\begin{equation}\label{eq:fijGW}
    f^{\rm GW}_{ij}\equiv - \frac{1}{8 \omega \omega_0} \left[(\omega^{+}_0)^2 \tilde{f}_{ij}(\omega^{+}_0 \Delta x_{ij}) - (\omega^{-}_0)^2 \tilde{f}_{ij}(\omega^{-}_0 \Delta x_{ij}) \right],
\end{equation}
and $\tilde{f}_{ij}(y) \equiv 4y^{-2} (1 - \cos y)
    - 2 {\rm sinc}(y),$
encodes the GW-induced dissipative coupling between $i$th and $j$th atoms. The total photon emission rate of the atomic array is $\Gamma_{\downarrow}(t) = \gamma_0 \sum_{i,j} F_{ij}(t) \langle \hat{\sigma}^{+}_{i} \hat{\sigma}^{-}_{j} \rangle (t),$
where $\langle \hat{\sigma}^{+}_{i} \hat{\sigma}^{-}_{j} \rangle (t) \equiv \Tr_{\rm A}\left(\hat{\sigma}^{+}_{i} \hat{\sigma}^{-}_{j} \hat{\rho}(t)\right)$, $\Tr_{\rm A}$ denotes trace over atoms, $\hat{\sigma}^{\pm}_j$ are the atomic raising and lowering operators for the $j$th atom, and $\hat{\rho}(t)$ is the density operator of the atomic system.

The collective photon emission from the array is characterized by the \textit{shape factor} $\mu$ of the array defined as: 
\begin{equation}\label{mu}
	\mu(t) \equiv \frac{1}{N^2} \sum_{i,j ; i \neq j} F_{ij}(t).
\end{equation}
In flat spacetime, all $N$ atoms can cooperate if the linear dimensions of the sample are smaller than the transition wavelength of each atom, leading to $\mu \to 1 - (1/N)$. In general, from $\abs{\mu} N$ one can construct a quantity that admits interpretation as the \textit{effective number of cooperating atoms}. For $\mu \to 0$, we get the incoherent emission rate of the array given as $\Gamma_{\downarrow}^{\rm inc}(t) = N \gamma_0 e^{-  \gamma_0 t}$.
The terms with $i=j$, for which $F_{ii} = f^{\rm M}_{ii} = 1$ and $f^{\rm GW}_{ii} = 0$, determine the contribution of individual atoms to emission rate. Note that the \textit{total} emission rate of a single atom remains unaffected by the GW up to first order in the GW amplitude~\cite{Qidong2020,Gray2021,Prokopec2023,Barman2023,Paczos2026}. The collective effects in the array are determined by the terms with $i \neq j$.  For $i \neq j$, $F_{ij}$ has a flat-spacetime contribution $f_{ij}^{\rm M}$ and a GW-induced contribution $h_{+} \cos(\omega t) f^{\rm GW}_{ij}$.  In general, for $i \neq j$, $f^{\rm GW}_{ij} \neq 0$. This means that although a single atom's total emission rate does not respond at first order in the amplitude of the GW (the leading-order effect scales as $h^2_{+}$~\cite{Gray2021}), the collective effects in the array's emission rate are sensitive to the GW. Following the identification of Minkowskian and GW contributions to $F_{ij}$, from Eq.~\eqref{mu} we write $\mu = \mu_{\rm M} + \mu_{\rm GW}$. Thus the flat-spacetime cooperation among the atoms is quantified by $\mu_{\rm M} N$ and, as we will see shortly, the GW-induced cooperation among the atoms can be quantified through $\mu_{\rm GW}(0)N/h_{+}$.

Based on the form of the dissipative coupling, we can identify three contributions to the photon emission rate of the array. (i) \textit{Incoherent Minkowskian emission} stems from each atom behaving independently of the others. This process involves incoherent photon emission at the atom's transition frequency $\omega_0$. This contribution is unaffected by the GW. (ii) \textit{Collective Minkowskian emission} arises from the collective behavior of the atoms but is insensitive to the GW. This process involves coherent photon emission at frequency $\omega_0$.  This component underlies the flat-spacetime quantum optical superradiance and may lead to a superradiant burst. In 1D arrays, the superradiant burst survives only for interatomic separations $d \lesssim 0.25 \lambda_0$~(see Fig.~\ref{fig:gwmink-kernels}(a) and refs.~\cite{Masson2020,Masson2022}), where $\lambda_{0}$ is the transition wavelength of each atom. (iii) \textit{GW-induced collective emission} results from GW-induced dissipative coupling between the atoms and is sensitive to the incident GW at first order in $h_{+}$. In contrast to Minkowskian collective emission, it involves coherent photon emission at shifted frequencies $\abs{\omega_0 \pm \omega}$, and is the contribution of interest to us. Crucially, it requires placing the atoms at $d \approx \lambda_0$ as shown below. Next, we discuss the distinction between flat spacetime and the GW-induced superradiance,  and identify the regimes where each arises.
%%%%%%%%%%%%%%%%%%%%%%%%%%%%%%%%%%%%%%%%%%%%%%%%%%%%%%%%%%%%%%%%
\paragraph*{Regimes of flat-spacetime and GW-induced superradiance---}
The condition required for the flat-spacetime and GW-induced cooperation of the atoms is different as can be seen from the forms of $f^{\rm M}_{ij}$ and $f^{\rm GW}_{ij}$. The two functions are compared in Figs.~\ref{fig:gwmink-kernels}(c-e). As $f^{\rm M}_{ij}$ attains a maximum for $\omega_0 \Delta x_{ij} \to 0$ and falls off as $(\omega_0 \Delta x_{ij})^{-1}$ away from the maximum, the flat-spacetime cooperation among atoms is strong only for interatomic spacing $d \ll \lambda_0$. As we are concerned with optical atomic transition frequencies and GW frequencies such that $\bar{\omega} \equiv \omega/\omega_0 \ll 1$, it is instructive to expand $f^{\rm GW}_{ij}$ as:
\begin{equation}\label{expansion1}
f^{\rm GW}_{ij}= \frac{1}{\bar{\omega}} \left(\bar{\omega} \tilde{f}_{ij}^{(1)}(\omega_0 \Delta x_{ij}) + \bar{\omega}^3 \tilde{f}_{ij}^{(3)}(\omega_0 \Delta x_{ij}) + ...\right),
\end{equation}
where $\tilde{f}_{ij}^{(1)}(y) \equiv  [\cos(y) - {\rm sinc}(y)]/2$
and $\tilde{f}_{ij}^{(3)}(y) \equiv - (y^2/12) [ {\rm sinc}(y) +  \cos(y)].$

From the above expansion, note that for $\sqrt{2 \pi^2/3} (\Delta x_{ij}/\lambda_{\rm gw}) \ll 1$ (i.e., approximately $\Delta x_{ij} \ll \lambda_{\rm gw}/2$), behavior of $f^{\rm GW}_{ij}$ is governed by $\tilde{f}_{ij}^{(1)}$ which, due to $\cos(\omega_0 \Delta x_{ij})$, maximizes at $\omega_0 \Delta x_{ij} = 2 \pi p, p \in \mathbb{Z}$, and enables dissipative coupling between distant atoms [see Figs.\,\ref{fig:gwmink-kernels}(c) and (e)]. In the sense of inducing an all-to-all dissipative coupling between atoms, the GW shapes collective dissipation of atoms in a way similar to that of a cavity or waveguide~\cite{Goban2015,Solano2017}, with the strength of dissipative coupling to this fictitious cavity or waveguide being proportional to the amplitude of the gravitational wave. Figure\,\ref{fig:gwmink-kernels}(e) shows that the GW-induced dissipative coupling between any two atoms in an array with $d=\lambda_0$ stays constant and prominent as long as the separation between the atoms is less than $\approx \lambda_{\rm gw}/2$, dips sharply as their separation approaches $\approx \lambda_{\rm gw}/2$, and remains small thereafter.

The distinction between the functional forms of $f^{\rm M}_{ij}$ and $f^{\rm GW}_{ij}$ allows for choosing an atom distribution that leads to stronger GW-induced collective dissipation and highly suppressed flat-spacetime collective dissipation: as the zeros of ${\rm sinc}(\omega_0 \Delta x_{ij})$ occur at $\omega_0 \Delta x_{ij} = p \pi, ~p \in \mathbb{Z} - \{0\}$, we consider the following spatial distribution of $N$ atoms (let $N$ be an odd number)
\begin{equation}\label{distribution}
	x_p = pd,~ p \in \left\{- \frac{N-1}{2},\frac{N-1}{2}\right\};
\end{equation}
so that we have $\omega_0 \Delta x_{ij} = 2 \pi (i-j)d/\lambda_0$.
%%%%%%%%%%%%%%%%%%%%%%%%%%%%%%%%%%%%%%%%%%%%%%%%%%
%%%%%%%%%%%%%%%%%%%%%%%%%%%%%%%%%%%%%%%%%%%%%%%
%%%%%%%%%%%%%%%%%%%%%%%%%%%%%%%%%%%%%%%%%%%%%%%%%%%%%%%%%%%%%%%%%%%%%%%
\begin{figure*}
	\centering
    \includegraphics[width=0.99\linewidth]{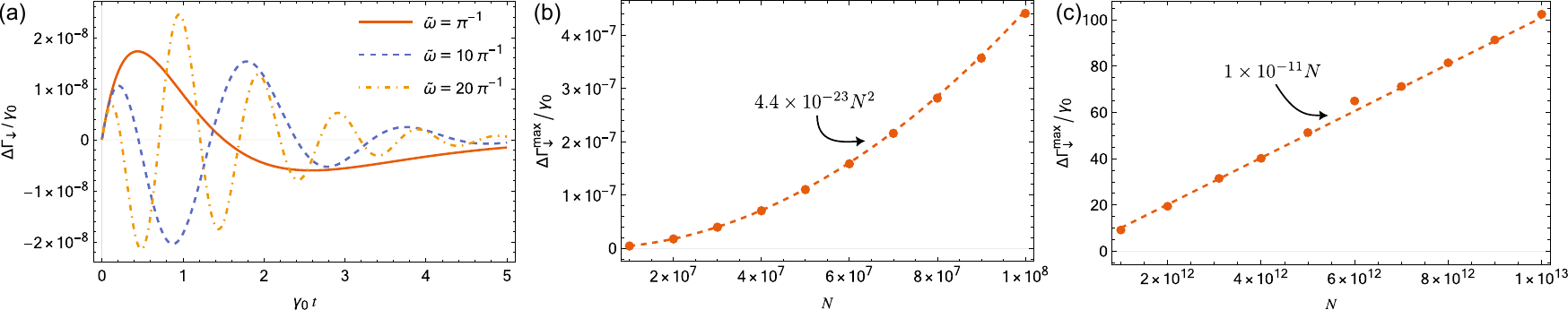}
	\caption{\textbf{(a)} The GW-induced correction $\Delta \Gamma_{\downarrow}$, in units of the spontaneous emission rate $\gamma_0$ of a single atom, versus $\gamma_0 t$, for $N = 10^7$ and different values of the ratio $\tilde{\omega} \equiv \omega/\gamma_0$. For higher values of $\tilde{\omega}$, the temporal profile of $\Delta \Gamma_{\downarrow}$ beats at the GW frequency [Eq.\,\eqref{decayrate4}]. \textbf{(b)} For total number of atoms in the array less than $\eta_{\rm max}$, $\Delta \Gamma_{\downarrow}$ scales quadratically with $N$. \textbf{(c)} For $N > \eta_{\rm max}$, $\Delta \Gamma_{\downarrow}$ scales at most linearly with $N$. All plots are for $h_+ = 10^{-21}$ and $\bar{\omega} = 10^{-12}$, and for an array with $d/\lambda_0 =1$ and $\theta_0 = 0.8 \pi$, which corresponds to $ 90 \% $ initial excitation probability of each atom in the array.}
	\label{fig:gwisr}
\end{figure*}
%%%%%%%%%%%%%%%%%%%%%%%%%%%%%%%%%%%%%%%%%%%%%%%%%%%%%%%%%%%%%%%%%%%%%%%%%%%%%%%
Notably, for $d/\lambda_{0} = 1$ the flat-spacetime dissipative coupling is zero, while the GW-induced dissipative coupling is maximized. The suppression of flat-spacetime collective dissipation for $d=\lambda_0$ is not an artifact of the scalar-light model as shown in our discussion of the full EM model~\cite{SM}.
Since the atom distribution prescribed in Eq.~\eqref{distribution} ensures that the Minkowskian cooperation among the atoms vanishes, we can quantify the effect of the GW on emission rate of the array through the quantity $\Delta \Gamma_{\downarrow} \equiv \Gamma_{\downarrow} - \Gamma^{\rm inc}_{\downarrow},$
which gives the GW-induced correction to the emission rate of the array and is obtained as\,\cite{SM}:
\begin{multline}\label{decayrate4}
	\frac{\Delta \Gamma_{\downarrow}(t)}{\gamma_0} = - \frac{ \mu_{\rm GW}(0) N^2 e^{-2 \gamma_0 t}}{4  \left(1 + \tilde{\omega}^2\right)} 
	\Bigg[-  \xi_0^2 e^{\gamma_0 t} \\
    +\xi_0^2  \left\{\left(2 + \tilde{\omega}^2\right) \cos( \tilde{\omega} \gamma_0 t) - \tilde{\omega}  \sin ( \tilde{\omega} \gamma_0 t )\right\} \\
	- 2 \xi_0 e^{\gamma_0 t} \left(1 + \tilde{\omega}^2\right) \left\{ \frac{\sin (\tilde{\omega} \gamma_0 t)}{\tilde{\omega}} - \cos ( \tilde{\omega} \gamma_0 t) \right\} \Bigg],
\end{multline}
where $\tilde{\omega} \equiv \omega/\gamma_0$ and $\xi_0 \equiv \cos\theta_0 - 1$ refers to the initial state of the array 
$\hat{\rho}(0) = \Pi_{i} \ket{\theta_{0},\varphi_{0}}_i  \leftindex_i {\bra{\theta_{0},\varphi_{0}}},~ \theta_{0} < \pi$. Here,
$\ket{\theta_{0},\varphi_{0}}_i = \sin(\theta_0/2) e^{-i \varphi_0/2} \ket{e}_{i} + \cos(\theta_0/2) e^{i \varphi_0/2} \ket{g}_{i}$,
with $\ket{e}$ and $\ket{g}$ denoting the excited and ground states, respectively, of a two-level atom.

The GW-imprint $\Delta \Gamma_{\downarrow}$ is nonzero only if the \textit{effective number of atoms showing GW-induced cooperation}
\begin{equation}\label{eta}
    \eta \equiv \frac{\abs{\mu_{\rm GW}(0)}N}{h_{+}} = \frac{1}{N} \abs{ \sum_{i,j;i \neq j} f^{\rm GW}_{ij}},
\end{equation}
is nonzero. Note that $\abs{\mu_{\rm GW}(0)}/h_+$ is independent of $h_+$ since $\mu_{\rm GW}(0)$ itself is proportional to $h_+$. We define the combination $\abs{\mu_{\rm GW}(0)} N/h_{+}$, and not just $\abs{\mu_{\rm GW}(0)} N$, as the effective number of atoms showing GW-induced cooperation because for $d=\lambda_0$ the number of atoms dissipatively coupled by the GW depends on the separation between the atoms and the ratio $\bar{\omega} \equiv \omega/\omega_0$, but is independent of $h_{+}$ [see Appendix\,\ref{apSec:crossover} and Figs.\,\ref{fig:gwmink-kernels}(c),(e)]. The factor of $\cos \omega t$ is omitted from this definition as it gives only a trivial overall modulation. In other words, $\eta$ is defined to reflect the long \textit{range} of the GW-induced dissipative coupling, rather than its strength (which is $\propto h_+$).

Figures~\ref{fig:gwmink-kernels}(a) and \ref{fig:gwmink-kernels}(b) show constant values of the flat-spacetime cooperation $\mu_{\rm M} N$ and GW-induced cooperation $\mu_{\rm GW}(0) N/h_{+}$, respectively, among the atoms as a function of the number of atoms and $d/\lambda_0$. Absolute values greater than unity of either quantity mean that there are atoms showing the corresponding cooperation. Positive (negative) value of either quantity mean that the corresponding collective effects encourage (suppress) photon emission\textemdash possibly causing superradiance (subradiance). From Figs.~\ref{fig:gwmink-kernels}(a) and \ref{fig:gwmink-kernels}(b), one notes that the flat-spacetime cooperation among the atoms is strong for $d/\lambda_0 \ll 0.25$ whereas the GW-induced cooperation predominates for $d/\lambda_0 \approx 1$, clearly demarcating different regimes of flat-spacetime and GW-induced cooperation. 

From Eq.~\eqref{decayrate4}, we find that if $\eta$ scales linearly with the total number of atoms $N$ in the array, $\Delta \Gamma_{\downarrow}(t)$ would scale quadratically with $N$. Indeed, we find that $\eta$ scales linearly with $N$ as long as the total length of the array is less than $\approx \lambda_{\rm gw}/2$, leading to $\Delta \Gamma_{\downarrow}(t) \propto h_+ N^2 \gamma_0$ as shown in Fig.\,\ref{fig:gwisr}(b). Once the total length of the array exceeds $\approx \lambda_{\rm gw}/2$, $\eta$ no longer scales with $N$ and gives $\Delta \Gamma_{\downarrow}(t) \propto h_+ N \gamma_0$ as shown in Fig.\,\ref{fig:gwisr}(c) [see Appendix\,\ref{apSec:crossover} for more details].
Moreover, since the time-derivative, $\Delta \dot{\Gamma}_{\downarrow}(0) = \mu_{\rm GW}(0) N^2 \xi_0 (3 \xi_0 + 4)\gamma^2_0/4$, of the GW-induced correction at $t=0$ is positive for $-2 < \xi_0 < -4/3$ (i.e., each atom initially occupying its excited state with a probability $> 67\%$) and $\mu_{\rm GW}(0) > 0$, the GW-induced correction to the emission rate shows a local maximum with a non-zero delay time [Fig.\,\ref{fig:gwisr}(a)]. In general, the GW-induced superradiant delay time will depend on $\xi_0$ and the ratio $\tilde{\omega} = \omega/\gamma_0$.

In analogy to flat-spacetime superradiance, the delay time and $N^2$ scaling can be used to mark the occurrence of \textit{GW-induced photon superradiance} during episodes of $\Delta \Gamma_{\downarrow}(t)>0$, see Figs.\,\ref{fig:gwisr}(a) and \ref{fig:gwisr}(b).
Note that for an array of fully excited atoms ($\theta_0 \to \pi$, i.e., $\xi_0 = -2$), $\Delta \Gamma_{\downarrow}(0) = 0$, meaning that such an array becomes sensitive to the gravitational wave only after sufficient quantum coherence, seeded by the vacuum fluctuations of the EM field, develops among the atoms. This vacuum-mediated GW-induced collective dissipation of atoms is a joint effect of general relativity and many-body quantum optics. We discuss the observability of this effect in Supplemental Material~\cite{SM}.

Interestingly, the temporal profile of the GW-induced superradiance beats at the GW frequency, with the beat profile encoding the GW phase [Fig.\,\ref{fig:gwisr}(a)]. This behavior is similar to the phenomenon of superradiance beats observed in flat spacetime under various settings like interference between two collective modes, multilevel atoms, or a mixture of two ensembles of two-level atoms having slightly different transition frequencies~\cite{Gross1982}. These beatings have been observed at nanoseconds to ultrafast timescales~\cite{Vrehen1977,Norcia2016,Han2021,Ariunbold2022}.  The beat profile of the GW-induced superradiance can in principle be used to ascertain the phase of the GW, for example, through a heterodyne detection of the superradiant field~\cite{Norcia2016,Liedl2024}. Further, a comparison of how a GW affects two co-located arrays oriented differently with respect to one another in the $xy$-plane can be used to determine its polarization~\cite{SM}.

In the GW-induced superradiance regime, the linewidth is still set by $\gamma_0$ as long as $Nh_+ < 1$ as shown in Appendix\,\ref{apSec:linewidth}. Moreover, the GW-induced dissipative coupling is robust against common noise sources in atom arrays, such as position disorder and partial filling (see Appendix~\ref{apSec: noise}). 
Thus, in relation to the feasibility of resolving general relativistic effects in the quantum domain using superradiance, we find the following key result: since we can separate the flat-spacetime and GW-induced total superradiance regimes, the extremely weak effects of general relativistic gravity on quantum systems can in principle be selectively amplified using the collective response of quantum emitters. Most importantly, while these intrinsically weak effects benefit from the characteristics of the superradiance phenomenon\textemdash such as amplification scaling faster than the system size\textemdash the weakness of the gravitational effect itself ensures that the superradiant linewidth broadening does not become a limiting factor that could impede resolution of GW-induced sidebands or prevent a sufficiently large number of atoms from cooperating (Appendix\,\ref{apSec:linewidth}). 
%%%%%%%%%%%%%%%%%%%%%%%%%%%%%%%%%%%%%%%%%%%%%%%%%%%%%%%%%%%%%%%%
%%%%%
%%%%%%%%%%%%%%%%%%%%%%%%%%%%%%%%%%%%%%%%%%%%%%%%%%%
% \section{Conclusion}
\paragraph*{Conclusion---}
We have demonstrated that engineered many-body quantum systems can selectively amplify weak spacetime-curvature effects.
% establishing a coherent many-body mechanism that amplifies the coupling between quantum emitters and gravitational waves. 
The weakness of gravitational effects has hindered empirical access to phenomena at the interface of general relativity and quantum mechanics. Our results show that the effects of spacetime geometry on collective emission can be isolated and amplified, with a scaling that surpasses linear scaling with system size.
An observation of the proposed vacuum-mediated GW-induced superradiance, even with modest precision in terms of GW parameters, would already constitute access to a previously untested physical regime of joint general-relativistic and quantum-mechanical effects. Thus, a high precision competitive with specialized GW detectors would not necessarily be the main objective of first experiments testing this effect. Beyond the GW context, the approach of selective amplification of gravitational effects through carefully engineered collective response of quantum systems may prove to be useful in the study of other weak effects at the interface of quantum theory and gravity~\cite{Howl2018,Guerreiro2020,Parikh2021a,Parikh2021b,Carney2024,Tobar2024}.
%%%%%%%%%%%%%%%%%%%%%%%%%%%%%%%%%%%%%%%%%%%%%%%%%%%%%%%%%%%%%%%%%%%%%%%%%%%%
%%%%%%%%%%%%%%%%%%%%%%%%%%%%%%%%%%%%%%%%%%%%%%%%%%%%%%%%%%%%%%%%%%%%%%%%%%%%%
\section*{Acknowledgments}
We thank Daniel Braun, Akhil Deswal, Swadheen Dubey, Sandeep K. Goyal, Kinjalk Lochan, Jorma Louko, Stuart J. Masson, Malte Schlosser, Jerzy Paczos, Germain Tobar, and Jun Ye for helpful discussions and useful comments. The authors acknowledge funding from Knut and Alice Wallenberg foundation through a Wallenberg Academy Fellowship No. 2021.0119.
%%%%%%%%%%%%%%%%%%%%%%%%%%%%%%%%%%%%%%%%%%%%%%%%%

\section*{Data availability}
The data that support the findings of this article are not publicly available. The data are available from the authors upon reasonable request.
%%%%%%%%%%%%%%%%%%%%%%%%%%%%%%%%%%%%%%%%%%%%%%%%%%%%%%%%%%%%%%%%%%%%%%%%%%%%%%%%%%%%%%%%%%%%% REFERENCES %%%%%%%%%%%%%%%

%%%%%%%%%%%%%%%%%%%%%%%%%%%%%%%%%%%%%%%%%%%%%%%%%%%%%%%%%%%%%%%%%
%%%%%%%%%%%%%%%% START OF APPENDIX %%%%%%%%%%%%%%%
% \newpage
% Figures, tables, equations and pages in the appendix are numbered A1, A2 etc.
\renewcommand{\thefigure}{A\arabic{figure}}
\renewcommand{\thetable}{A\arabic{table}}
\renewcommand{\theequation}{A\arabic{equation}}
\renewcommand{\thesection}{A\arabic{section}}
\setcounter{section}{0}
\setcounter{figure}{0}
\setcounter{table}{0}
\setcounter{equation}{0}
% References continue the numbering from the main text.
\setcounter{secnumdepth}{3}  % sections+subsections+subsubsections
\section*{Appendix}
\section{Crossover from GW-induced superradiance to GW-induced normal radiance}\label{apSec:crossover}
The range of the GW-induced dissipative coupling between $i$th and $j$th atoms is encoded in $f^{\rm GW}_{ij}$, and the net total for all atom pairs determines the $N$-scaling of $\eta$, and hence that of GW-induced emission rate. The GW-induced crosstalk among the atoms can be understood as follows: consider a reference atom $i_0$ and analyze its dissipative coupling to atoms on other sites $j$, i.e., $f^{\rm GW}_{i_0j}$. Figure~\ref{fig:gwmink-kernels}(c) does so for $i_0 = 0$. For $d=\lambda_0$, the reference atom is dissipatively coupled prominently to other atoms up to $j = j_{\rm max}$. Beyond that, $f^{\rm GW}_{i_0j}$ starts oscillating between negative and positive values, and remains small. Thus, $\sum_{j} f^{\rm GW}_{i_0 j}$ takes its maximum value for summation up to  $j = j_{\rm max}$.  

For a given $\bar{\omega}$, the $j_{\rm max}$ value can be estimated by considering the expansion of $\mu_{\rm GW}$ as a power series in $\bar{\omega} \ll 1$ using Eq.~\eqref{expansion1}. Note that the coefficients $\tilde{f}_{ij}^{(1)}$ and $\tilde{f}_{ij}^{(3)}$ are out-of-phase. For a given $\bar{\omega}$ and atom $i_{0}$, beyond a certain $ j = j_{\rm max}$, the  contribution of $\tilde{f}_{ij}^{(3)}$ becomes dominant as it scales as $(\Delta x_{ij})^2$. Therefore, for the distribution of atoms prescribed in Eq.~\eqref{distribution}, we can estimate $j_{\rm max}$ for $i_0 = 0$ simply by solving $\tilde{f}_{0j_{\rm max}}^{(1)} + \bar{\omega}^2 \tilde{f}_{0j_{\rm max}}^{(3)} = 0$. We get $j_{\rm max} \approx \sqrt{3/2\pi^2} \bar{\omega}^{-1}$, which is slightly below the actual value of $j_{\rm max} \approx \sqrt{3/2\pi^2} (5/4\bar{\omega}) \approx 1/2\bar{\omega}$ deduced from Figs.~\ref{fig:gwmink-kernels}(c),(e) because for this simple estimate we have not taken into account the higher order terms in $\bar{\omega}$. As the discussion following Eq.~\eqref{expansion1} suggests, the same condition follows if one intuitively demands that for atoms in an array with $d = \lambda_0$ to show GW-induced cooperation, the length of the array must be less than half of the GW wavelength.
These observations determine $\eta_{\rm max}$, the maximum number of atoms that show GW-induced cooperation. For example, for $\bar{\omega} = 10^{-12}$, we get $\eta_{\rm max} \sim 10^{11}$.

We have used the $\eta_{\rm max}$ estimate to guide us in plotting $\Delta \Gamma_{\downarrow}/\gamma_0$ as a function of $N$ in Figs.~\ref{fig:gwisr}(b) and \ref{fig:gwisr}(c). The figures show the $N$-scaling characteristics of $\Delta \Gamma^{\rm max}_{\downarrow}/\gamma_0$. As long as the total number of atoms in the array is less than $j_{\max}$, $\eta$ scales linearly with $N$ up to a maximum value $\eta_{\rm max}$.  Consequently, $\Delta \Gamma^{\rm max}_{\downarrow}$ scales quadratically with $N$. Once $N$ exceeds $\eta_{\rm max}$, the sensitivity increases at most linearly if more atoms are added to the array and $\Delta \Gamma^{\rm max}_{\downarrow}$ scales only linearly with $N$. This marks the crossover from the GW-induced superradiance to GW-induced normal radiance. Note that GW-induced normal radiance is also of collective origin. 
%%%%%%%%%%%%%%%%%%%%%%%%%%%%%%%%%%%%%%%%%%%%%%%%%%%%%%%%%
%%%%%%%%%%%%%%%%%%%%%%%%%%%%%%%%%%%%%%%%%%%%%%%%%
\begin{figure*}
	\centering
	\includegraphics[width=0.99\linewidth]{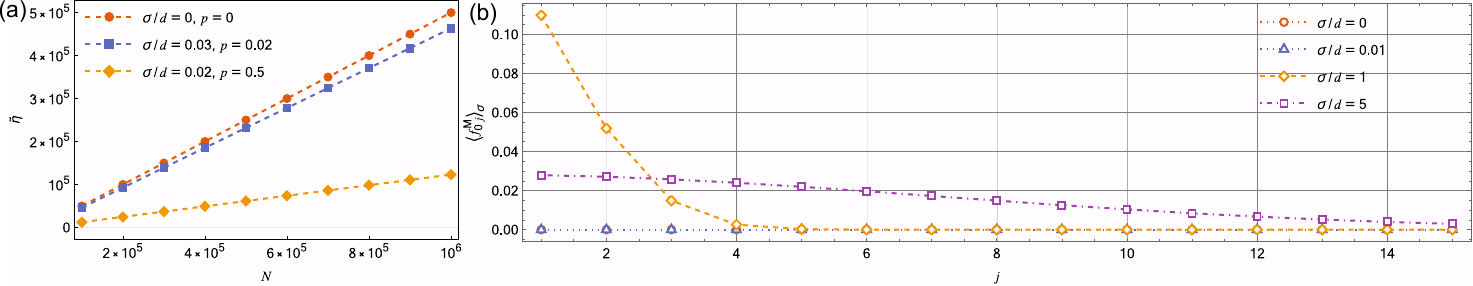}
	\caption{\textbf{(a)} Impact of position disorder and missing atoms on GW-induced cooperation. The scaling of effective number of atoms showing GW-induced cooperation as a function of the total number of atoms $N$, under the effect of classical position disorder and partial filling in the array. We model position disorder by assuming that each atom's position obeys a normal distribution with variance $\sigma$ and that each atom can be missing from its site with a probability $0 \leq p \leq 1$. We compare the ideal case $(\sigma/d =0, p=0)$ with two recent realizations of coherent atom arrays with parameters $(\sigma/d = 0.03, p = 0.02)$~\cite{Norcia2024} and $(\sigma/d = 0.02, p = 0.5)$~\cite{Manetsch2025}. The linear scaling of $\tilde{\eta}$ with $N$ remains robust even in the presence of position disorder in present-day optical lattices and under partial filling with $p \lesssim 0.1$. \textbf{(b)} Impact of position disorder on flat-spacetime cooperation among atoms. The plotted quantity is $\langle f^{\rm M}_{0j}\rangle_{\sigma} \equiv \int_{- \infty}^{\infty} \dd{x}_i q(x_i) \int_{- \infty}^{\infty} \dd{x}_j q(x_j) f^{\rm M}_{0j}(x_i - x_j)$. The red circles and blue triangles overlap with each other for all practical purposes. A position disorder of $\sigma/d \lesssim 0.01$ around the prescribed atomic positions does not revive the flat-spacetime dissipative coupling between the atoms. Such values of $\sigma/d$ are achievable in present-day ordered arrays~\cite{Norcia2024,Manetsch2025}.}
	\label{fig:mungw-noisy}
\end{figure*}
%%%%%%%%%%%%%%%%%%%%%%%%%%%%%%%%%%%%%%%%%%%%%%%%%
\section{Linewidth and retardation effects in the GW-induced superradiance regime}\label{apSec:linewidth}
For an array with interatomic spacing approximately equal to the atomic transition wavelength, the flat-spacetime total cooperation, $\mu_{\rm M} N$, among the atoms vanishes, prohibiting flat-spacetime superradiance [see Supplemental Material~\cite{SM} for more details on the full EM model]. The separation between flat-spacetime and GW-induced superradiance regimes has significant consequences for the characteristic time period over which atoms decay in the two regimes. As can be seen from Eq.~\eqref{decayrate4}, the characteristic time scale for photon emission in the GW-induced superradiance regime ($d=\lambda_0$, $\mu_{\rm M}=0$) is still set by $1/\gamma_0$ and the corresponding linewidth is $\Delta = \gamma_0/2 \pi$.
On the other hand, in the flat-spacetime superradiance regime i.e., for $d \ll \lambda_0/4$ and large $N$, characteristic time scale for photon emission (also known as the superradiance time) is given as $T = 1/\gamma_0 (\mu_{\rm M} N + 1)$~\cite{Rehler1971,Agarwal1970a,SM}. In this case, one would have $\mu_{\rm M} \approx 1$ and consequently would obtain $\Delta_{\rm flat} \approx \gamma_0(1 + N)/2\pi$, which gives the superradiantly broadened linewidth of $\Delta_{\rm flat} \approx \gamma_0 N/2\pi$~\cite{Rehler1971,Gross1982,Malz2022}.

Consequently, in the GW-induced superradiance regime the number of atoms that can be expected to exhibit GW-induced cooperation before retardation and non-Markovian effects become important are determined by the photon coherence length set by $\gamma_0$ i.e., $l_{\rm c} = c/\gamma_0$. Therefore, the condition to ensure validity of the Markovian approximation in the GW-induced superradiance regime is $N d \ll c/\gamma_0$, which gives the critical number of atoms to be $N_{\rm critical, gw} = \omega_0/(2 \pi \gamma_0)$, and requires $N \ll N_{\rm critical, gw}$.
For example, for the narrow spin-forbidden transition $(\leftindex^1 S_0 - \leftindex^3 P_1)$ in barium-138 having a spontaneous emission rate $\gamma_0 \sim 2\pi \times 25~\mathrm{kHz}$ and wavelength $791~\mathrm{nm}$ (superradiance has been observed on this transition~\cite{Kim2018}), one obtains $N_{\rm critical, gw} \sim 10^{9}$. For a longer lived transition, the critical number of atoms would be even higher.
On the other hand, the condition for the Markovian approximation to hold in the flat-spacetime superradiance regime is $N d \ll c/N \gamma_0$~\cite{Lehmberg1970I,Lehmberg1970II}, giving the critical number of atoms as $N_{\rm critical, flat} = \sqrt{\omega_0/2 \pi \gamma_0}$.
For the above-mentioned transition one obtains $N_{\rm critical, flat} \sim 10^4$. Therefore, for our purposes, the number of atoms that can be dissipatively coupled by the GW are not prohibitively limited by retardation and the related non-Markovian effects.

One can still quantify the retardation and ensuing non-Markovian effects in the GW-induced superradiance regime, in a 1D array of total length $L$, in terms of $\chi = L\gamma_0/c$, i.e., the total length of the array normalized by the photon coherence length $\approx c/\gamma_0$~\cite{KSinha2020}. Wherever we consider a specific number of atoms and atomic linewidth in our analysis, we ensure $\chi \ll 1$, meaning that retardation effects and the associated non-Markovian effects remain subleading. As $\chi$ approaches one, the retardation effects become important~\cite{KSinha2020,Windt2025}.
%%%%%%%%%%%%%%%%%%%%%%%%%%%%%%%%%%%%%%%%%%%%%%%%%
\section{GW-induced dissipative coupling under position disorder and partial filling}\label{apSec: noise}
Here, we analyze the robustness of GW-induced dissipative coupling to position disorder and partial filling in an ordered atom array. For modeling these effects we may use the observation that atoms in an optical lattice, rather than taking precise positions as prescribed in Eq.~\eqref{distribution}, undergo zero-point motion $\Delta x_{\rm ZPF} = \sqrt{\hbar/2m\Omega}$, where $m$ is the mass of the atom and $\Omega$ is the trap frequency. Moreover, atoms can be missing from some sites.  We include the impact of classical position disorder and missing atoms on the scaling of $\Delta \Gamma_{\downarrow}$ with $N$ by studying their impact on the scaling of $\eta$ [see Eq.~\eqref{decayrate4}]. We assume that each atom's position follows a normal distribution $q(x_i)$ of variance $\sigma$ around its unperturbed position. The impact of missing atoms (partial filling) can be included by replacing $f^{\rm GW}_{ij}$ in $\eta$ by $f^{\rm GW}_{ij;{\rm ideal}} - f^{\rm GW}_{ij;{\rm holes}}$, where 
\begin{equation}
    f^{\rm GW}_{ij;{\rm holes}} = \begin{cases}
        & f^{\rm GW}_{ij;{\rm ideal}},~ \text{either $i$ or $j$ missing} \\
        & 0,~ \text{otherwise}
    \end{cases}.
\end{equation}
If each atom can be missing with a probability $0 \leq p \leq 1$, then the effective number of atoms showing GW-induced cooperation under position disorder and partial filling is given by the absolute value (which we will denote by $\tilde{\eta}$) of
\begin{equation}\label{mugw2}
	- e^{- 4 \pi^2 (\sigma/\lambda_0)^2} \frac{(1-p)^2}{8 \omega \omega_0 N} \sum_{i,j ; i \neq j} f^{\rm GW}_{ij}\left(\omega^{\pm}_0 \Delta x_{ij} \right).
\end{equation}
In Fig.~\ref{fig:mungw-noisy}(a), we plot $\tilde{\eta}$ for the ideal case and for two different values of the set $(\sigma/d,p)$. For $\sigma/d$ and $p$, we chose values $(0.03, 0.02)$ and $(0.02, 0.5)$ based on recent works realizing coherent arrays with $10^3$~\cite{Norcia2024} and $6 \times 10^3$~\cite{Manetsch2025} atoms, respectively. Figure~\ref{fig:mungw-noisy}(b) shows that the flat-spacetime cooperation between atoms does not revive under a position disorder of $\sigma/d \lesssim 0.01$.
%%%%%%%%%%%%%%%%%%%%%%%%%%%%%%%%%%%%%%%%%%%%%%%%%%%
%%%%%%%%%%%%%%%%%%%%%%%%%%%%%%%%%%%%%%%%%%%%%%%%%%%
%%%%%%%%%%%%%%%% SUPPLEMENT LIST %%%%%%%%%%%%%%%

\clearpage

%%%%%%%%%%%%%%%% START OF SUPPLEMENT %%%%%%%%%%%%%%%

% Figures, tables, equations and pages in the supplement are numbered S1, S2 etc.
\renewcommand{\thefigure}{S\arabic{figure}}
\renewcommand{\thetable}{S\arabic{table}}
\renewcommand{\theequation}{S\arabic{equation}}
\renewcommand{\thepage}{S\arabic{page}}
\renewcommand{\thesection}{S\arabic{section}}
\setcounter{section}{0}
\setcounter{figure}{0}
\setcounter{table}{0}
\setcounter{equation}{0}
\setcounter{page}{1} % not 0 as \newpage already started a supplementary page
% References continue the numbering from the main text.
\setcounter{secnumdepth}{3}  % sections+subsections+subsubsections

\clearpage
\onecolumngrid

%%%%%%%%%%%%%%%% SUPPLEMENTARY TEXT %%%%%%%%%%%%%%%
\begin{center}
	\section*{Supplemental Material for ``Gravitational Wave-Induced Superradiance in Ordered Atomic Arrays"}
    Navdeep Arya$^1$\orcidlink{https://orcid.org/0000-0003-0730-4835} and
	Magdalena Zych$^{1,2}$\orcidlink{0000-0002-8356-7613}\\
    $^1$Department of Physics, Stockholm University, SE-106 91 Stockholm, Sweden\\
    $^2$ARC Centre for Engineered Quantum Systems, School of Mathematics and Physics, The University of Queensland, St. Lucia, Queensland, 4072, Australia
\end{center}
%%%%%%%%%%%%%%%%%%%%%%%%%%%%%%%%%%%%%%%%%%%%%%%%%%%%%%%%%%%%%%%%%%%%%%%%%%%%%%%%%
\section{Quantization on a GW-background Spacetime}\label{apSec: quantization}
Assuming that the gravitational field is weak, we can decompose the metric into the flat Minkowski metric $\eta_{\mu \nu}$ plus a small perturbation $h_{\mu \nu}$: $g_{\mu \nu} = \eta_{\mu \nu} + h_{\mu \nu}, ~ \abs{h_{\mu\nu}} \ll 1$.
The solution of the vacuum linearized Einstein equations in the Transverse-Traceless gauge leads to the following line element for a plane GW spacetime (with $z$ as the GW propagation direction)~\cite{Maggiore2007}
\begin{equation}\label{line_element1}
	\dd{s^2} = - \dd{t^2} + \dd{z^2} + \left\{1 + h_+ \cos[\omega(t-z)]\right\} \dd{x^2}
	+ \left\{1 - h_+ \cos[\omega(t-z)]\right\} \dd{y^2} + 2 h_{\times} \cos[\omega (t - z)] \dd{x} \dd{y},
\end{equation}
where $h_+$ and $h_{\times}$ are the amplitudes of the ``plus'' and ``cross'' polarizations, respectively, and $\omega$ is the frequency of the gravitational wave.

For our analysis we take the background to be a plane gravitational wave (GW) with $+$ polarization.
In terms of the light-cone coordinates $(u, v, x, y)$, the line element then takes the form $\dd{s^2} = - \dd{u} \dd{v} + g_{ab} \dd{x^a} \dd{x^b}$,
where $a \in \{x,y\}$, $u \equiv t - z$, $v \equiv t + z$, $g_{ab}=\text{diag}(f(u)^2 ,g(u)^2)$, $f(u) \equiv \sqrt{1 + h_+ \cos(\omega u)}$, and $g(u) \equiv \sqrt{1 - h_+ \cos(\omega u)}$.
A complete orthonormal set of the solutions of  Klein-Gordon equation in this spacetime is obtained as~\cite{Garriga1991}
\begin{equation}\label{mode_fxns}
	\mathcal{U}_{k_v,k_a}(u,v,x,y) = \frac{1}{(2\pi)^{3/2}\sqrt{2k_v}} \left(\det(g_{ab}(u))\right)^{-1/4} \exp\left(- \frac{i}{4 k_v} \int_{0}^{u} \dd{u'}  g^{ab}(u') k_a k_b \right) e^{ik_a x^a - i k_v v},
\end{equation}
where the modes are normalized as $(\mathcal{U}_{k_v,k_a},\mathcal{U}_{k'_v,k'_a}) = \delta^2(k^a - k'^a) \delta(k_v - k'_v)$. The integral in Eq.~\eqref{mode_fxns} evaluates to $\int_{0}^{u} \dd{u'}  g^{ab}(u') k_a k_b = (k_x^2 + k_y^2)u - (h_+/\omega) (k_x^2 - k_y^2) \sin(\omega u)$.

Next, we quantize the field $\Phi$ by imposing the usual commutation relations between $\Phi$ and its conjugate momentum $\pi_{\Phi}$ and obtain~\cite{Garriga1991}
\begin{equation}
	\hat{\Phi}(u,v,x^a) = \int \dd^2{k}_a \dd{k_v} \left(\hat{a}_{k_v,k_a} \mathcal{U}_{k_v,k_a} + \hat{a}^{\dagger}_{k_v,k_a} \mathcal{U}^*_{k_v,k_a}\right),
\end{equation}
with $	[\hat{a}_{k_v,k_a} , \hat{a}^{\dagger}_{k'_v,k'_a}] = \delta^2(k_a - k'_a) \delta(k_v - k'_v)$.\\
%%%%%%%%%%%%%%%%%%%%%%%%%%%%%%%%%%%%%%%%%%%%%%%%%%%%%%%%%%%%%%%%%%%%%%%%%%%%%%%%%

With the atom-field interaction Hamiltonian specified in the following section, the open system dynamics of the atom array is governed by the two-point Wightman function of the field $\hat{\Phi}$. To first order in $h_+$, it is given as~\cite{Qidong2020,Gray2021}
\begin{equation}\label{wightman}
	\begin{split}
		W(\tilde{x}_i,\tilde{x}'_j) &= \expval{\hat{\Phi}(\tilde{x}_i) \hat{\phi}(\tilde{x}'_j)}{0}  = \frac{1}{2(2\pi)^{3}} \int \dd^2{k}_a \int \frac{\dd{k_v}}{k_v} \exp\left(- \frac{i}{4 k_v} (k_x^2 + k_y^2) \Delta u_{ij} \right) e^{ik_a \Delta r^a_{ij} - i k_v \Delta v_{ij}}  \\
		&+ \frac{ih_+}{4 (2\pi)^3 \omega} \int \dd^2{k}_a \int \frac{\dd{k_v}}{k^2_v}  \exp\left(- \frac{i}{4 k_v} (k_x^2 + k_y^2) \Delta u_{ij} \right) e^{ik_a \Delta r^a_{ij} - i k_v \Delta v_{ij}}  (k_x^2 - k_y^2) \\
		&\hspace{6cm} \times \cos\left(\omega\frac{u_i + u'_j}{2}\right) \sin\left(\omega\frac{\Delta u_{ij}}{2}\right)\\
		& \equiv W_{\mathrm{M}}(\tilde{x}_i,\tilde{x}'_j) + W_{\mathrm{GW}}(\tilde{x}_i,\tilde{x}'_j),
	\end{split}
\end{equation}
where $\tilde{x}$ denotes a spacetime event, $\Delta r^{a}_{ij}$ is the separation between the $i$th and $j$th atoms in the $xy$-plane, and $W_{\mathrm{M}}(\tilde{x}_i,\tilde{x}'_j)$ and $W_{\mathrm{GW}}(\tilde{x}_i,\tilde{x}'_j)$ are, respectively, the Minkowski and gravitational wave contributions to the Wightman function up to first order in $h_{+}$. 

We are interested in the response of an atomic array arranged transverse to the direction of propagation of the GW. In this connection, we note a crucial modification in the contribution of transverse field modes $\vb{k}_{\perp}$ to the response of the array due to presence of the gravitational wave. To this end, we perform the transverse angular integration in the Minkowskian and GW contributions to the Wightman function by writing $\dd^2{k_a} = \dd{k_{\perp}} k_{\perp} \dd{\phi_{\vb{k}_{\perp}}}$,  $k_x = k_{\perp} \cos\phi_{\vb{k}_{\perp}}$, and $k_y = k_{\perp} \sin\phi_{\vb{k}_{\perp}}$.
The $\phi_{\vb{k}_{\perp}}$-integral in $W_{\mathrm{M}}(\tilde{x}_i,\tilde{x}'_j)$ evaluates to: (a) $2 \pi J_0(\sqrt{2} l k_{\perp})$ for $\Delta x_{ij} = \Delta y_{ij} = l$, and (b) $2\pi J_0(l k_{\perp})$ for either $\Delta x_{ij} = l, \Delta y_{ij} = 0$ or $\Delta x_{ij} = 0,\Delta y_{ij} = l$.
Further, the $\phi_{\vb{k}_{\perp}}$-integral in $W_{\mathrm{GW}}(\tilde{x}_i,\tilde{x}'_j)$ vanishes for $\Delta x_{ij} = \Delta y_{ij}$ but evaluates to: (a) $-2\pi J_2(l k_{\perp})$ for $\Delta x_{ij} = l, \Delta y_{ij} = 0$ and (b)  $2\pi J_2(l k_{\perp})$ for $\Delta x_{ij} = 0,\Delta y_{ij} = l$.
We thus learn that the contribution of transverse field modes $k_{\perp}$ to the response of atoms in a 1D array along $\hat{x}$-direction in the Minkowski space is governed by $J_0(k_{\perp} \Delta x_{ij})$, but in the presence of a GW the contribution sensitive to the GW is governed by $J_2(k_{\perp} \Delta x_{ij})$.
%%%%%%%%%%%%%%%%%%%%%%%%%%%%%%%%%%%%%%%%%%%%%%%%%%%%%%%%%%%%%%%%%%%%%%%%%%%%%%%%%%%%%%%%
\section{The Superradiant Master Equation}
The coupling of atoms to a given polarization of an electromagnetic field, under the dipole-approximation, can be modeled by an interaction Hamiltonian $\hat{H}_{\rm I} = g \sum_i \hat{\mathfrak{m}}_i(t) \hat{\Phi}(t,\vb{r}_i)$. Here, 
the sum is over the atoms, $\vb{r}_i$ is the position of $i$th atom, and $g$ is a coupling constant assumed to be small~\cite{Gross1982,Breuer2002}. Further, $\hat{\mathfrak{m}}_j \equiv (i/2) (\hat{\sigma}^{-}_j - \hat{\sigma}^{+}_j)$, where $\hat{\sigma}^{\pm}_j$ are the atomic raising and lowering operators for the $j$th atom, and $\hat{\Phi}(t,\vb{r})$ is a real massless quantum scalar field. This interaction Hamiltonian is known to be a good model of atom-light interaction when no exchange of angular momentum is involved~\cite{Martinez2013}. In addition, the pointlike interaction Hamiltonian prescription leads to transition probabilities invariant under arbitrary diffeomorphisms~\cite{Martinez2020,Martinez2021}. To cleanly identify the key features of interest, we first work within the scalar-light model as laid out above and consider the full electromagnetic model in section~\ref{SMsec:EM model} below. The free Hamiltonian of each atom is $\hat{H}_{i} = \omega_0 \hat{\sigma}^{z}_{i}/2$, where $\hat{\sigma}^{z}_{i}$ is the Pauli $z$-matrix for the $i$th atom and $\omega_0$ is the transition frequency. For the initial state of the array, we consider the product state~\cite{allen1975}
\begin{equation}\label{eq:initial_state}
	\hat{\rho}(0) = \Pi_{i} \ket{\theta_{0},\varphi_{0}}_i  \leftindex_i {\bra{\theta_{0},\varphi_{0}}},~ \theta_{0} < \pi,
\end{equation}
where $\ket{\theta_{0},\varphi_{0}}_i = \sin(\theta_0/2) e^{-i \varphi_0/2} \ket{e}_{i} + \cos(\theta_0/2) e^{i \varphi_0/2} \ket{g}_{i}$,
with $\ket{e}$ and $\ket{g}$ denoting the excited and ground states, respectively, of a two-level atom.

For the atom-field composite system under consideration, the superradiant master equation takes the form~\cite{Gross1982}
\begin{multline}\label{SR-ME-scalar}
	\dv{\hat{\rho}(t)}{t} = - i \Big[\sum_j \hat{H}_j,\hat{\rho}(t)\Big]  - \frac{i g^2}{2} \sum_{i,j} \int_{0}^{t} \dd{s} W(\tilde{x}_i,\tilde{x}_j) \\
	\times \Big\{ \left(\hat{\mathfrak{m}}_i \hat{\sigma}^{-}_j \hat{\rho}(t) - \hat{\sigma}^{-}_j \hat{\rho}(t)\hat{\mathfrak{m}}_i\right) e^{i\omega_0 s} - \left(\hat{\mathfrak{m}}_i \hat{\sigma}^{+}_j \hat{\rho}(t) - \hat{\sigma}^{+}_j \hat{\rho}(t) \hat{\mathfrak{m}}_i\right) e^{- i\omega_0 s}\Big\}
	+ \text{h.c. of the integrals},
\end{multline}
where $\hat{\rho}(t)$ is the density operator of the atomic system, $W(\tilde{x}_i,\tilde{x}_j) \equiv \expval{\hat{\Phi}(t,\vb{r}_i) \hat{\Phi}(t-s ,\vb{r}_j)}{0}$ is the two-point Wightman function of the field, and `h.c.' denotes hermitian conjugate.
For the light-cone coordinates $u \equiv t-z$ and $v \equiv t+z$, we have $\Delta u_{ij} = t_i - z_i - t_j + z_j$ and $\Delta v_{ij} = t_i + z_i - t_j - z_j$. However, we consider an array of atoms in the $xy$-plane, that is, in the plane transverse to GW propagation direction. Therefore, $z$-coordinate of all the atoms is the same. Thus, $\Delta u_{ij} = t_i  - t_j$ and $\Delta v_{ij} = t_i - t_j$. For $t_i = t, t_j = t - s$, we get $\Delta u_{ij} = s, \Delta v_{ij} = s$, and $u_i + u_j = 2t-s$.
Further, as noted in the previous section, since GW contribution to Wightman function vanishes for any pair of atoms for which $\Delta x_{ij} = \Delta y_{ij}$, we assume that the atoms are separated only along the $ \hat{x} $-direction.  
%%%%%%%%%%%%%%%%%%%%%%%%%%%%%%%%%%%%%%%%%%%%%%%%%%%%%%%%%%%%%%%
\subsection{Minkowskian Contribution to the Master Equation}\label{apSec: Mink contri}
The Minkowskian contribution to the Wightman function can be written as
\begin{equation}
	\expval{\hat{\Phi}(t,\vb{r}_i) \hat{\Phi}(t-s ,\vb{r}_j)}{0}_{\rm M} = \frac{1}{2(2\pi)^{2}} \int_{0}^{\infty} \dd{k_{\perp}} k_{\perp} J_0(k_{\perp} \Delta x_{ij}) \int \frac{\dd{k_v}}{k_v} \exp\left\{-i\left(\frac{k^2_{\perp}}{4 k_v} + k_v\right) s \right\}.
\end{equation}
We change the integration variables from $(k_{\perp},k_v)$ to $(k_{\perp},k_z)$ using transformations
\begin{equation}\label{coord-trans}
	k \equiv k_v + \frac{k^2_{\perp}}{4k_v},  k_z \equiv k_v - \frac{k^2_{\perp}}{4k_v},
\end{equation}
with $\dd{k}_v \dd{k}_{\perp}/k_v = \dd{k}_z \dd{k}_{\perp}/k$, and a further change of integration variable from $k_z$ to $k$ using $\dd{k}_z/k = \dd{k}/k_z$ leads to
\begin{equation}
	\expval{\hat{\Phi}(t,\vb{r}_i) \hat{\Phi}(t-s ,\vb{r}_j)}{0}_{\rm M}  = \frac{1}{(2\pi)^{2}} \int_{0}^{\infty} \dd{k_{\perp} k_{\perp}} J_0(k_{\perp} \Delta x_{ij}) \int_{0}^{\infty} \frac{\dd{k}}{k_z} e^{-i k s}.
\end{equation}
Substituting the expression for the Minkowskian Wightman function in the superradiant master equation and raising the upper limit of the $s$-integral to infinity under the Markov approximation, the Minkowskian contribution (marked by subscript $\rm M$) to the evolution of the density operator of the atomic array is given as
\begin{multline}\label{rho_M}
	\left[\dv{\hat{\rho}(t)}{t}\right]_{\rm M} = - \frac{i g^2}{2(2\pi)^2} \sum_{i,j} \int_{0}^{\infty} \dd{s} \int_{0}^{\infty} \dd{k}_{\perp} k_{\perp} J_0(k_{\perp} \Delta x_{ij}) \int_{0}^{\infty} \dd{k} \frac{ \Theta(k-k_{\perp})}{\sqrt{k^2 - k^2_{\perp}}} e^{-i k s} \\
	\times \Big\{ \left(\hat{\mathfrak{m}}_i \hat{\sigma}^{-}_j \hat{\rho}(t) - \hat{\sigma}^{-}_j \hat{\rho}(t) \hat{\mathfrak{m}}_i\right) e^{i\omega_0 s} - \left(\hat{\mathfrak{m}}_i \hat{\sigma}^{+}_j \hat{\rho}(t) - \hat{\sigma}^{+}_j \hat{\rho}(t) \hat{\mathfrak{m}}_i\right) e^{- i\omega_0 s}\Big\} + \text{h.c. of the integrals}.
\end{multline}
The terms coming from $\left(\hat{\mathfrak{m}}_i \hat{\sigma}^{+}_j \hat{\rho}(t) - \hat{\sigma}^{+}_j \hat{\rho}(t) \hat{\mathfrak{m}}_i\right) e^{- i\omega_0 s}$ and $\Im \int_{0}^{\infty} \dd{s} \exp{-i(k - \omega_0)s}$ in above equation contribute only to the Lamb shift and Van der Waals dipole-dipole interaction among the atoms. We focus on the dissipative dynamics of the atomic array (see Sec.\,\ref{SMSec:dephasing-cdd}) and drop these terms to obtain
\begin{equation}
	\left[\dv{\hat{\rho}(t)}{t}\right]_{\rm M,DD} = - \frac{ i g^2 \pi}{2(2\pi)^2} \sum_{i,j} \int_{0}^{\infty} \dd{k}_{\perp} k_{\perp} J_0(k_{\perp} \Delta x_{ij})  \frac{ \Theta(\omega_0 - k_{\perp})}{\sqrt{\omega_0^2 - k^2_{\perp}}} \left(\hat{\mathfrak{m}}_i \hat{\sigma}^{-}_j \hat{\rho}(t) - \hat{\sigma}^{-}_j \hat{\rho}(t) \hat{\mathfrak{m}}_i\right)  + \text{h.c.},
\end{equation}
for the Minkowskian contribution to the dissipative dynamics. The subscript $\rm DD$ in $\left[\dv*{\hat{\rho}(t)}{t}\right]_{\rm M,DD}$, and hereafter, denotes `dissipative dynamics'.
Further, evaluating the $k_{\perp}$-integral and dropping the fast rotating terms under rotating-wave approximation by replacing $\hat{\mathfrak{m}}_i \hat{\sigma}^{-}_j \hat{\rho}(t)$ and $\hat{\sigma}^{-}_j \hat{\rho}(t) \hat{\mathfrak{m}}_i$ by $(-i/2)\hat{\sigma}^{+}_i \hat{\sigma}^{-}_j \hat{\rho}(t)$ and $(-i/2) \hat{\sigma}^{-}_j \hat{\rho}(t) \hat{\sigma}^{+}_i$, respectively, and incorporating the h.c. terms we obtain
\begin{equation}\label{MinkDD1}
	\left[\dv{\hat{\rho}(t)}{t}\right]_{\rm M,DD} = \frac{\pi g^2 \omega_0}{2(2\pi)^2} \sum_{i,j} \frac{\sin(\omega_0 \Delta x_{ij})}{\omega_0\Delta x_{ij}}\left(- \frac{1}{2} \acomm{\hat{\sigma}_i^+ \hat{\sigma}^{-}_j}{\hat{\rho}(t)} + \hat{\sigma}^{-}_j \hat{\rho}(t) \hat{\sigma}^{+}_i\right).
\end{equation}
%%%%%%%%%%%%%%%%%%%%%%%%%%%%%%%%%%%%%%%%%%%%%%%%%%%%%%%%%%%%%%%%%%%%%%%%%%%%%%%%%%%%%%%%%%%%%
\subsection{GW contribution to the Master Equation}\label{apSec: GW contri}
The part of the Wightman function sensitive to the presence of the GW can be cast as
\begin{multline}\label{gw-wightman}
	W_{\mathrm{GW}}(\vb{r}_i,s;\vb{r}_j,t-s) = -\frac{h_+}{16 (2\pi)^2 \omega} \left[e^{- i \omega  t} (e^{i \omega s} - 1)-  e^{i \omega  t} (e^{- i \omega s} - 1)\right]  \\ \times \int_{0}^{\infty} \dd{k}_{\perp} k^3_{\perp} \int_0^{\infty} \frac{\dd{k_v}}{k^2_v}  \exp\left\{-i \left(\frac{k^2_{\perp} }{4 k_v} + k_v\right)s \right\} J_2(k_{\perp}\Delta x_{ij}).
\end{multline}
Next, we make a change of integration variables from $(k_{\perp}, k_v)$ to $(k_{\perp}, k)$ in above equation with transformations given in Eq.~\eqref{coord-trans}. To this end, we use two different representations of Bessel-K function. First, from the representation~\cite{Gradshteyn}
\begin{equation}
	K_{\nu}(xz) = \frac{z^{\nu}}{2} \int_{0}^{\infty} \dd{t} t^{-\nu-1} \exp[-\frac{x}{2} \left(t + \frac{z^2}{t}\right)],
\end{equation}
for $\abs{\arg{z}}<\pi/4$ or $\abs{\arg{z}}=\pi/4$ and $\Re{\nu}<1$, we obtain
\begin{equation}\label{rep1}
	k_{\perp} K_{1}(i k_{\perp} s) = \frac{k^2_{\perp}}{4} \int_{0}^{\infty} \frac{\dd{k_v}}{k^2_v}  \exp[-i s \left(k_v + \frac{k_{\perp}^2}{4 k_v}\right)].
\end{equation}
From another representation of the Bessel-K~\cite{Gradshteyn}:
\begin{equation}
	u K_{1}(u \mu) = \int_{u}^{\infty} \dd{x} \frac{x e^{-\mu x}}{\sqrt{x^2 - u^2}},
\end{equation}
for $u > 0, \Re(\mu) > 0$, we can write
\begin{equation}\label{rep2}
	k_{\perp} K_1(i k_{\perp } s ) = \int_{k_{\perp}}^{\infty} \dd{k} e^{-i k s } \frac{k}{\sqrt{k^2 - k^2_{\perp}}}. 
\end{equation}
Using representations~\eqref{rep1} and \eqref{rep2}, we achieve the desired transformation and write
\begin{multline}\label{gw-wightman1}
	W_{\mathrm{GW}}(\vb{r}_i,t;\vb{r}_j,t-s) = - \frac{h_+}{8 (2\pi)^2 \omega} \left[e^{- i \omega  t} (e^{i \omega s} - 1)-  e^{i \omega  t} (e^{- i \omega s} - 1)\right]  \\ \times \int_{0}^{\infty} \dd{k}_{\perp} k_{\perp} \int_{0}^{\infty} \dd{k} \frac{2k}{\sqrt{k^2 - k^2_{\perp}}} \Theta(k - k_{\perp})  e^{-i k s} J_2(k_{\perp}\Delta x_{ij}).
\end{multline}
Using Eq.~\eqref{gw-wightman1} in the superradiant master equation [Eq.~\eqref{SR-ME-scalar}], we find that the contribution of the GW to the evolution of the array is governed by (after rotating-wave approximation)
\begin{equation}
	\begin{split}
		&\left[\dv{\hat{\rho}(t)}{t}\right]_{\rm GW} =  \frac{g^2 h_+}{32 (2\pi)^2 \hbar^2 \omega} \sum_{i,j} \int_{0}^{t} \dd{s}  \int_{0}^{\infty} \dd{k}_{\perp} k_{\perp} \int_{0}^{\infty} \dd{k}  \frac{2k}{\sqrt{k^2 - k^2_{\perp}}} \Theta(k - k_{\perp}) J_2(k_{\perp}\Delta x_{ij})\\
		&\times \Big\{ \left(\hat{\sigma}^{+}_i \hat{\sigma}^{-}_j \hat{\rho}(t) - \hat{\sigma}^{-}_j \hat{\rho}(t) \hat{\sigma}^{+}_i\right) \left[e^{- i \omega  t} (e^{- i (k - \omega^{+}_0) s} - e^{- i (k - \omega_0) s})-  e^{i \omega  t} (e^{- i (k - \omega^{-}_0) s} - e^{-i (k- \omega_0) s})\right]  \\
		&+ \left(\hat{\sigma}^{-}_i \hat{\sigma}^{+}_j \hat{\rho}(t) - \hat{\sigma}^{+}_j \hat{\rho}(t) \hat{\sigma}^{-}_i\right) \left[e^{- i \omega  t} (e^{- i (k + \omega^{-}_0) s} - e^{- i (k + \omega_0) s})-  e^{i \omega  t} (e^{- i (k + \omega^{+}_0) s} - e^{- i (k + \omega_0) s})\right] \Big\} + \text{h.c.},
	\end{split}
\end{equation}
where $\omega^{\pm}_0 \equiv \omega_0 \pm \omega$.
For an array of total length $L$, the bath correlation time $t_{\rm B}$ is of the order of $L/c$. If the upper limit $t$ of the $s$-integral is already large enough as compared to $t_{\rm B} \sim L/c$, we can safely raise the upper limit of the $s$-integral to infinity, this is known as the \textit{Markovian approximation}. Raising the upper limit of the $s$-integral to infinity we get an equation that describes the dynamics for $t \gg t_{\rm B}$:
\begin{equation}\label{MEGW-ll}
	\begin{split}
		&\left[\dv{\hat{\rho}(t)}{t}\right]_{\rm GW} =  \frac{g^2 h_+}{32 (2\pi)^2 \hbar^2 \omega} \sum_{i,j} \int_{0}^{\infty} \dd{s}  \int_{0}^{\infty} \dd{k}_{\perp} k_{\perp} \int_{0}^{\infty} \dd{k}  \frac{2k}{\sqrt{k^2 - k^2_{\perp}}} \Theta(k - k_{\perp}) J_2(k_{\perp}\Delta x_{ij})\\
		&\times \Big\{ \left(\hat{\sigma}^{+}_i \hat{\sigma}^{-}_j \hat{\rho}(t) - \hat{\sigma}^{-}_j \hat{\rho}(t) \hat{\sigma}^{+}_i\right) \left[e^{- i \omega  t} (e^{- i (k - \omega^{+}_0) s} - e^{- i (k - \omega_0) s})-  e^{i \omega  t} (e^{- i (k - \omega^{-}_0) s} - e^{-i (k- \omega_0) s})\right]  \\
		&+ \left(\hat{\sigma}^{-}_i \hat{\sigma}^{+}_j \hat{\rho}(t) - \hat{\sigma}^{+}_j \hat{\rho}(t) \hat{\sigma}^{-}_i\right) \left[e^{- i \omega  t} (e^{- i (k + \omega^{-}_0) s} - e^{- i (k + \omega_0) s})-  e^{i \omega  t} (e^{- i (k + \omega^{+}_0) s} - e^{- i (k + \omega_0) s})\right] \Big\} + \text{h.c.}
	\end{split}
\end{equation}

As discussed in Sec.\,\ref{SMSec:dephasing-cdd} below, we focus on the dissipative dynamics of the atomic array. From the $s$-integral in Eq.~\eqref{MEGW-ll}, we get both the Dirac delta function (resonant) and Cauchy principal value (off-resonant) terms contributing to the dissipative dynamics of the array under the influence of the GW. The dominant contribution comes from the resonant terms, moreover the contribution stemming from the Cauchy principal value terms scales very weakly with the number of atoms. Consequently, the dominant GW-induced contribution to the dissipative dynamics of the array is obtained as
\begin{multline}
	\left[\dv{\hat{\rho}(t)}{t}\right]_{\rm GW,DD} \approx  \frac{\pi g^2 h_+ \cos\omega t}{32 (2\pi)^2 \hbar^2 \omega} \sum_{i,j}  \int_{0}^{\infty} \dd{k}_{\perp} k_{\perp} \int_{0}^{\infty} \dd{k}  \frac{2k}{\sqrt{k^2 - k^2_{\perp}}} \Theta(k - k_{\perp}) J_2(k_{\perp}\Delta x_{ij})\\
	\times \Big\{ \left(\hat{\sigma}^{+}_i \hat{\sigma}^{-}_j \hat{\rho}(t) - \hat{\sigma}^{-}_j \hat{\rho}(t) \hat{\sigma}^{+}_i\right) \left[\delta(k - \omega^{+}_0)-   \delta(k - \omega^{-}_0) \right]  + \text{h.c.}
\end{multline}
%%%%%%%%%%%%%%%%%%%%%%%%%%%%%%%%%%%%%%%%%%%%%%%%%%
Performing the $k$-integral and evaluating the $k_{\perp}$-integral as
\begin{equation}
\begin{split}
	\int_{0}^{\infty} \dd{k}_{\perp} k_{\perp} \frac{2 a }{\sqrt{a^2 - k^2_{\perp}}}  \Theta(a - k_{\perp}) J_2(k_{\perp}\Delta x_{ij}) &= a^2 \frac{4 -4 \cos (a \Delta x_{ij}) -2 (a \Delta x_{ij}) \sin (a \Delta x_{ij})}{(a\Delta x_{ij})^2} \\
    &\equiv a^2 \tilde{f}_{ij}(a\Delta x_{ij}),
    \end{split}
\end{equation}
one obtains
\begin{equation}
		\left[\dv{\hat{\rho}(t)}{t}\right]_{\rm GW,DD} = \frac{\pi g^2 h_{+} \cos \omega t}{32 (2\pi)^2 \hbar^2 \omega} \sum_{i,j}  \Big[(\omega^{+}_0)^2 \tilde{f}_{ij}(\omega^{+}_0 \Delta x_{ij}) - (\omega^{-}_0)^2 \tilde{f}_{ij}(\omega^{-}_0 \Delta x_{ij}) \Big]   \left(\hat{\sigma}^{+}_i \hat{\sigma }^{-}_j \hat{\rho}(t) - \hat{\sigma}^{-}_j \hat{\rho}(t) \hat{\sigma}^{+}_i\right) + \text{h.c.}
\end{equation}
Finally, incorporating the h.c. terms, the total (including both the Minkowskian and GW contributions) dissipative dynamics of the density operator $\hat{\rho}(t)$ is governed by:
\begin{equation}\label{TotalDD}
	\left[\dv{\hat{\rho}(t)}{t}\right]_{\rm DD} = \gamma_0 \sum_{i,j} F_{ij}(t)  \left(- \frac{1}{2} \acomm{\hat{\sigma}_i^+ \hat{\sigma}^{-}_j}{\hat{\rho}(t)} + \hat{\sigma}^{-}_j \hat{\rho}(t) \hat{\sigma}^{+}_i\right),
\end{equation}
where $F_{ij}(t)$ is as defined in Eq.~\eqref{Fij}.
%%%%%%%%%%%%%%%%%%%%%%%%%%%%%%%%%%%%%%%%%%%%%%%%%%%%%%%%%%%%%%%%%%%%%%%%%%%%%%%%%%%%%%%%%%%%%%%%%%%%%%%%%%%%%
\section{Emission rate of the atomic array}\label{apSec:decayrate}
\noindent
From Eq.~\eqref{TotalDD} the total emission rate of the atomic array is obtained as
\begin{equation}\label{decayrate}
	\Gamma_{\downarrow}(t) = \gamma_0 \sum_{i,j} F_{ij}(t) \langle \hat{\sigma}^{+}_{i} \hat{\sigma}^{-}_{j} \rangle (t),
\end{equation}
where $\langle \hat{\sigma}^{+}_{i} \hat{\sigma}^{-}_{j} \rangle (t) \equiv \Tr_{\rm A}\left(\hat{\sigma}^{+}_{i} \hat{\sigma}^{-}_{j} \hat{\rho}(t)\right)$.
To obtain the decay rate as a function of time, we need to evaluate $\langle \hat{\sigma}^{+}_{i} \hat{\sigma}^{-}_{j} \rangle (t)$. 
We start by noting
\begin{equation}\label{iden}
	\sum_{i,j} F_{ij}(t) \expval{\hat{\sigma}^{+}_{i} \hat{\sigma}^{-}_{j}} = \sum_{i,j ; i \neq j} F_{ij}(t) \expval{\va{\hat{\sigma}}_{i} \cdot \va{\hat{\sigma}}_{j} - \hat{\sigma}^{z}_{i} \hat{\sigma}^{z}_{j}} + \sum_{j} F_{jj}(t) \expval{\hat{\sigma}^{+}_{j} \hat{\sigma}^{-}_{j}}.
\end{equation}
%%%%%%%%%%%%%%%%%%%%%%%%%%%%%%%%%%%%%%%%%%%%%%%%%%
Recalling that the permutation operator, the operator corresponding to the interchange of $i$th and $j$th atom, is given by $\hat{P}_{ij} = \left(\hat{\mathds{1}} + \va{\hat{\sigma}}_{i} \cdot \va{\hat{\sigma}}_{j}\right)/2$,
we write
\begin{equation}\label{iden1}
	\sum_{i,j} F_{ij} \expval{\hat{\sigma}^{+}_{i} \hat{\sigma}^{-}_{j}} =\frac{1}{4} \sum_{i,j ; i \neq j} F_{ij} \expval{2 \hat{P}_{ij} - \hat{\mathds{1}}} - \frac{1}{4} \sum_{i,j ; i \neq j} F_{ij} \expval{\hat{\sigma}^{z}_{i} \hat{\sigma}^{z}_{j}} + \sum_{j} F_{jj} \expval{\hat{\sigma}^{+}_{j} \hat{\sigma}^{-}_{j}}.
\end{equation}
%%%%%%%%%%%%%%%%%%%%%%%%%%%%%%%%%%%%%%%%%%%%%%%%%%%%%
To obtain the emission rate of the array as a function of time for large $N$ case, we introduce an approximation to decouple the multi-particle mean values. For the initial state of the atomic array prescribed in Eq.~\eqref{eq:initial_state}, an appropriate approximation is to assume~\cite{Agarwal1970a,Rehler1971,Agarwal1971c}
\begin{equation}\label{approx2}
    \langle \hat{\sigma}^z_i \hat{\sigma}^z_{j} \rangle = \langle \hat{\sigma}^z_i \rangle \langle \hat{\sigma}^z_{j} \rangle,~ (i \neq j).
\end{equation}
Note that the approximation does not necessarily imply that the density operator of the atomic array can be written as direct product of density operators of individual atoms. By decoupling mean values at a higher order, one finds that corrections to Eq.~\eqref{approx2} are of the order $1/N$~\cite{Agarwal1971c}. Therefore, the inaccuracy introduced by the approximation is quite small for large values of $N$, the case of interest to us. Under this approximation, we obtain
\begin{equation}
	\sum_{i,j} F_{ij} \ev*{\hat{\sigma}^{+}_{i} \hat{\sigma}^{-}_{j}} =\frac{1}{4}  \sum_{i,j ; i \neq j} F_{ij} \ev{2 \hat{P}_{ij} - \hat{\mathds{1}}} - \frac{1}{4} \sum_{i,j ; i \neq j} F_{ij} \expval{\hat{\sigma}^z_i} \expval{\hat{\sigma}^z_{j}} +  \sum_{j} \ev*{\hat{\sigma}^{+}_{j} \hat{\sigma}^{-}_{j}}.
\end{equation} 
Also, using $\hat{\sigma}_j^+ \hat{\sigma}^{-}_j = (\hat{\mathds{1}} + \hat{\sigma}^{z}_j)/2$,
we obtain $\sum_{j} \ev*{\hat{\sigma}^{+}_{j} \hat{\sigma}^{-}_{j}} = \frac{1}{2} \sum_{j}  \expval{(\hat{\mathds{1}} + \hat{\sigma}^{z}_j)} = \frac{1}{2}  \Big( N + \sum_j \ev*{\hat{\sigma}^{z}_j} \Big)$.
Since $F_{ij}$ is symmetric under the exchange of $i$ and $j$, the evolution governed by Eq.~\eqref{TotalDD} preserves the permutation symmetry if the atomic array is initialized in a permutationally symmetric state. As $\ev*{\hat{\sigma}^z_j}$ is the same for all atoms then, we can write
\begin{equation}
	\begin{split}
		\sum_{i,j} F_{ij} \expval{\hat{\sigma}^{+}_{i} \hat{\sigma}^{-}_{j}} &= \frac{1}{4} \sum_{i,j ; i \neq j} F_{ij} \expval{2 \hat{P}_{ij} - \hat{\mathds{1}}} - \frac{\expval{\hat{\sigma}^z_a} \expval{\hat{\sigma}^z_{b}}}{4} \sum_{i,j ; i \neq j} F_{ij} + \frac{1}{2}  \Big( N + \sum_j \ev{\hat{\sigma}^{z}_j} \Big) \\
		&= \frac{1}{4} \sum_{i,j ; i \neq j} F_{ij} \expval{2 \hat{P}_{ij} - \hat{\mathds{1}}} - \frac{\left(\sum_{a=1}^{N}\expval{\hat{\sigma}^z_a} \right)^2}{4 N^2} \sum_{i,j ; i \neq j} F_{ij} + \frac{1}{2}  \Big( N + \sum_j \ev{\hat{\sigma}^{z}_j} \Big).
	\end{split}
\end{equation} 
If the atomic array is initialized in a permutationally symmetric state then $\langle \hat{P}_{ij} \rangle = \mathds{1}$. Therefore,
\begin{equation}
	\sum_{i,j} F_{ij} \expval{\hat{\sigma}^{+}_{i} \hat{\sigma}^{-}_{j}} 
	= \frac{1}{4} \left(1 - \frac{\left(\sum_{a=1}^{N}\expval{\hat{\sigma}^z_a} \right)^2}{N^2}\right) \sum_{i,j ; i \neq j} F_{ij} + \frac{1}{2} \Big( N + \sum_j \ev{\hat{\sigma}^{z}_j} \Big).
\end{equation} 
In terms of the total (dimensionless) energy of the atomic system: $W = \frac{1}{2} \sum_{l} \expval{\hat{\sigma}^z_l}$, we write
\begin{equation}
	\sum_{i,j} F_{ij} \expval{\hat{\sigma}^{+}_{i} \hat{\sigma}^{-}_{j}} = \frac{1}{4} \left(1 - \frac{4 W^2}{N^2}\right) \sum_{i,j ; i \neq j} F_{ij}(t) + \frac{1}{2} (N + 2 W).
\end{equation}
Using Eq.~\eqref{decayrate} we obtain a differential equation for $W(t)$:
\begin{equation}\label{dWdt2S}
	\dv{W(t)}{t} =   \gamma_0 \mu(t) \left(W(t) + \frac{N}{2}\right) \left(W(t) - \frac{N}{2} - \frac{1}{\mu(t)}\right),
\end{equation}
where $\mu$ is as defined in the main text.
%%%%%%%%%%%%%%%%%%%%%%%%%%%%%%%%%%%%%%%%%%%%%%%%%%
For a time-dependent $\mu$, Eq.~\eqref{dWdt2S} can still be solved analytically in the GW-induced superradiant regime (i.e, $d = \lambda_0$, such that $\mu_{\rm M} = 0$) with an ansatz of the form $W(t) = w(t) + \zeta(t)$,
where $w(t) = - N (1 + \xi_0 e^{-\gamma_0 t})/2$ is the flat-spacetime incoherent contribution to the total energy of the system and $\zeta(t)$ is the GW-induced contribution. It is reasonable to assume that $\zeta(t) \ll 1, \forall t$ (recall that $W(t)$ is the total \textit{dimensionless} energy of the atomic system). This assumption puts an upper limit on the total number of atoms $N$ in the array but the limit is so high that we anyway don't need to saturate it. Under the initial condition $\zeta(t=0)=0$ (because $w(0) = - N (1 + \xi_0)/2$ accounts for the total dimensionless energy of the atomic system at this moment), we obtain
\begin{equation}\label{eq:W-ansatz}
    \zeta(t)= \frac{ \mu_{\rm GW}(0)  N^2 \xi_0 e^{-2 \gamma_0 t}}{4  \left(1 + \tilde{\omega}^2\right)} \Big[ \left\{\xi_0 \tilde{\omega}^2 + 2 e^{\gamma_0 t} \left(1 + \tilde{\omega}^2 \right)  \right\} \frac{\sin(\tilde{\omega} \gamma_0 t)}{\tilde{\omega}} + \xi_0  e^{\gamma_0 t} - \xi_0  \cos( \tilde{\omega} \gamma_0 t) \Big],
\end{equation}
and consequently for the GW-induced correction to the total emission rate of the array we obtain Eq.\,\eqref{decayrate4} of the main text.

%%%%%%%%%%%%%%%%%%%%%%%%%%%%%%%%%%%%%%%%%%%%%%%%%%%%%%%%%%%%%%%%
\section{Comparison to the full electromagnetic model}\label{SMsec:EM model}

In the main text, we work in the scalar-light model for simplification, wherein the electromagnetic (EM) polarization effects are neglected. The description of flat-spacetime collective effects in the scalar-light model is recovered from the full EM model for a particular choice of the orientation of the dipole moments of atoms with respect to the length of the array, as shown below. As compared to the full EM model, the scalar-light model leaves out the so-called near-field terms in the dissipative coupling $F_{ij}$ between the atoms (and more generally in the Green function). For two atoms separated by $\Delta x_{ij}$, the near-field terms are proportional to either $(\omega_0 \Delta x_{ij})^{-2}$ or $(\omega_0 \Delta x_{ij})^{-3}$. Therefore, even for a general orientation of the atomic dipoles w.r.t.~interatomic separation vector, the near-field terms do not enable collective effects for interatomic spacing $d \approx \lambda_0$.

Concretely, the interaction Hamiltonian between the atoms and the electromagnetic field is given by $\hat{H}_I = - \sum_j \hat{d}_j^{\mu} \hat{E}_{\mu}(\tilde{x}_j)$, where $d_j^{\mu}$ is the electric dipole moment four-vector of the $j$th atom, $E_{\mu} \equiv F_{\mu \nu} u^{\nu}$, $F_{\mu\nu}$ is the electromagnetic field strength tensor, and $u^{\nu}$ is the four-velocity of an atom~\cite{Anandan2000}. The interaction Hamiltonian takes the form $\hat{H}_I = - \hat{d}_j \hat{E}^j(t,\vb{r}_j)$ in the rest frame of the atoms, i.e., for $u^{\mu} = (1,0,0,0)$.  In the interaction picture, the dipole moment operator $\hat{\va{d}}(t)$ is given in terms of its matrix elements as $\hat{\va{d}}(t) = \va{d}\, \hat{\hat{\sigma}}^{-} \exp(-i \omega_0 t) + \va{d}^* \hat{\hat{\sigma}}^{+} \exp(i \omega_0 t)$, 
where $\va{d} \equiv \bra{g} \hat{\va{d}}(t = 0) \ket{e}$. We consider all atomic dipoles to be aligned, i.e., $\va{d}_j = \va{d}$, and denote the unit dipole vector by $\vu{d}_{\rm a}$. In the following, we analyze the dissipative coupling between atoms in the flat and GW-background spacetimes with two specific goals: (a) to show that the suppression of flat-spacetime collective dissipation for $d=\lambda_0$ is not an artifact of the scalar-light model, and (b) to check that the GW-induced superradiance regime itself is not affected in the full EM model. 
%%%%%%%%%%%%%%%%%%%%%%%%%%%%%%%%%%%%%%%%%%%%%%%%%%%%%%%%%%%%%%%%%%%%%%%%%%%%%%%%%%%%%%%%%%%%
\subsection{Dissipative coupling rate in flat-spacetime}
The dissipative coupling rate between atoms in flat-spacetime, in the full EM model, is given as~\cite{Gross1982,Garcia2017,Carmichael2000}:
\begin{equation}\label{eq:EM-dcr}
    \Gamma^{\rm M}_{ij} = \gamma_0 f^{\rm M}_{ij} = \frac{3\gamma_0}{2} \Bigg\{ \frac{\sin(\omega_0 \Delta x_{ij})}{\omega_0 \Delta x_{ij}}(1 - (\vu{d}_{\rm a} \cdot \vu{x}_{ij})^2) + (1 - 3 (\vu{d}_{\rm a} \cdot \vu{x}_{ij})^2) \left[\frac{\cos(\omega_0 \Delta x_{ij})}{\omega^2_0 \Delta x^2_{ij}}-\frac{\sin(\omega_0 \Delta x_{ij})}{\omega^3_0 \Delta x^3_{ij}}\right]\Bigg\},
\end{equation}
where $\vu{x}_{ij}$ is the unit vector along the separation between the atoms. 
\begin{enumerate}
    \item \textbf{Recovering the scalar-light model:} If we chose the orientation of the atomic dipole moment vector $\vu{d}_{\rm a}$ such that
    \begin{equation}
        1 - 3 (\vu{d}_{\rm a} \cdot\vu{x}_{ij})^2 = 0,
    \end{equation}
    which corresponds to choosing the angle $\vartheta$ between the dipole vector and the separation between atoms such that $\cos^2 \vartheta_{\rm m} = 1/3$, we recover the dissipative coupling
    \begin{equation}
    f^{\rm M}_{ij} = \frac{\sin(\omega_0 \Delta x_{ij})}{\omega_0 \Delta x_{ij}},
\end{equation}
as in the scalar-light model. The angle $\vartheta_{\rm m}$ is referred to as the \textit{magic angle}.
%%%%%%%%%%%%%%%%%%%%%%%%%%%%%%%%%%%%%%%%%%%%%%%%%%%%%%%%%%%%%%%%%%%%%%%%%%%%%%%
\item \textbf{Considering general orientation of atomic dipoles:} In this case, both far- and near-field terms contribute in general. With our prescription [Eq.\,\eqref{distribution}] of atom distribution in a 1D array, the far-field contribution, ${\rm sinc}(\omega_0 \Delta x_{ij})$, vanishes and the contribution of the near-field terms is obtained from Eq.\,\eqref{eq:EM-dcr} as
\begin{equation}
    f^{\rm M}_{ij;{\rm near-field}} = \frac{3}{2} (1 - 3 (\vu{d}_{\rm a} \cdot\vu{x}_{ij})^2)\frac{1}{4 \pi^2 (i-j)^2}, i \neq j.
\end{equation}
The corresponding correction to \textit{effective number of atoms showing flat-spacetime cooperation} is given as
\begin{equation}
    (\mu_{\rm M} N)_{\rm near-field} = \frac{1}{N} \sum_{i,j; i \neq j} f^{\rm M}_{ij;{\rm near-field}} = (1 - 3 (\vu{d}_{\rm a} \cdot\vu{x}_{ij})^2) \frac{3}{8 \pi^2 N} \sum_{i,j; i \neq j}\frac{1}{(i-j)^2}.
\end{equation}

For a large number of atoms in the array ($N \gg 1$), we can approximate
\begin{equation}
    \frac{1}{N}\sum_{i,j; i \neq j}^{N}\frac{1}{(i-j)^2} \approx \frac{\pi^2}{3},
\end{equation}
and thus
\begin{equation}\label{eq:muN-approx1}
    (\mu_{\rm M} N)_{\rm near-field} \approx \frac{1}{8}(1 - 3 (\vu{d}_{\rm a} \cdot\vu{x}_{ij})^2).
\end{equation}
Crucially, $(\mu_{\rm M} N)_{\rm near-field}$ does not scale with $N$. Further, for flat-spacetime superradiant burst to occur (which we want to avoid as a bare minimum requirement), $\mu_{\rm M} N$ must be greater than 1. This can be seen from the expression of the radiated intensity in flat spacetime superradiance (obtained by solving Eq.\,\eqref{dWdt2S} in the absence of a GW)~\cite{Eberly1969,Agarwal1970a}:
    \begin{equation}
    		\Gamma_{\rm M}(t) = \frac{\gamma_0}{2 \mu_{\rm M}}(\mu_{\rm M} N + 1)^2 \sech^2\left(\frac{t-t_{\rm d}}{t_{\rm sr}}\right),
    \end{equation}
    where $t_{\rm sr}^{-1} \equiv \gamma_0 (\mu_{\rm M} N + 1)$ is known as the \textit{superradiance time} and
    \begin{equation}
        t_{\rm d} = \frac{\ln(\mu_{\rm M} N)}{\gamma_0 (\mu_{\rm M} N + 1)},
    \end{equation}
    is the \textit{superradiant delay time}\textemdash the time at which the photon emission rate would scale as $(\mu_{\rm M}N +1)^2$. Clearly, $t_{\rm d} > 0$ requires $\mu_{\rm M} N > 1$. This observation can also be expressed by saying that the superradiant burst requires dissipative coupling beyond nearest-neighbors, as analyzed in detail in ref.\cite{Mok2023}. This is so because $\mu_{\rm M} N$ itself has the interpretation of the effective number of atoms to which an atom dissipatively couples in a sample. For perfect cooperation, $\mu_{\rm M} \to 1 - \frac{1}{N}$, and therefore $\mu_{\rm M} N \to N - 1$. The maximum positive correction contributed by the near-field terms to the effective number of atoms showing flat-spacetime cooperation in Eq.~\eqref{eq:muN-approx1} is $1/8 \approx 0.125$, insufficient to revive the flat-spacetime superradiant burst.

    In contrast, for $d=\lambda_0$ and as long as the total length of the array is less than $\approx \lambda_{\rm gw}/2$, we found in the main text that $\mu_{\rm GW}(0) N$ scales linearly with $N$\textemdash meaning that all such atoms are dissipatively coupled due to the GW (though the strength of this coupling is weak as it is proportional to $h_+$). Consequently, the GW-induced correction to total photon emission rate scales as $\sim h_+ N^2$, see Eq.\,\eqref{decayrate4} and the discussion that follows. Thus even when $\vartheta \neq \vartheta_{\rm m}$, for $d=\lambda_0$ we have a regime in which the flat-spacetime dissipative coupling between atoms leads to a contribution to total emission rate scaling linearly with $N$ (the same scaling as that of incoherent photon emission) whereas the GW-induced dissipative coupling leads to a contribution that scales quadratically with $N$ and beats at the GW frequency. Also note that the ranges of the flat-spacetime and GW-induced dissipative couplings are drastically different: while the former is extremely short-ranged ($\sim \cos (\omega_0 \Delta x_{ij})/ (\omega_0 \Delta x_{ij})^2$), the latter is long-ranged ($\sim \cos (\omega_0 \Delta x_{ij})$) and ultimately only limited by the wavelength of the GW $\lambda_{\rm gw} \gg \lambda_0$. All these features underscore the qualitative prominence of the GW-induced collective effects in the $d \approx \lambda_0$ regime even when $\vartheta \neq \vartheta_{\rm m}$.
\end{enumerate}

%%%%%%%%%%%%%%%%%%%%%%%%%%%%%%%%%%%%%%%%%%%%%%%%%%%%%%%%%%%%%%%%%%%%%%%%%%%%%%%%%%%%%%%%%%%%
\subsection{Dissipative coupling rate induced by the GW}
Similar to the flat-spacetime case above, the full EM model adds near-field terms to the dissipative coupling $F_{ij}$ in addition to the far-field terms already captured by the scalar-light model on a GW-background spacetime. The dissipative coupling in the full EM model is computed from the gauge-invariant quantity $\hat{d}_i^{\mu}\ev{\hat{E}_{\mu}(\tilde{x}_i) \hat{E}_{\nu}(\tilde{x}_j)}{0} \hat{d}_j^{\nu}$ through essentially the same steps as for the scalar-light model, with the addition of the polarization factors.

As an alternative, we can take the more intuitive perspective of the local Lorentz coordinates of the array to readily check using Eq.~\eqref{eq:EM-dcr} itself that the GW-induced superradiance regime is not affected in the full EM model.
To this end, first note that the GW-induced superradiance is enabled by the leading-order (precisely, zeroth-order) term $\tilde{f}^{(1)}_{ij}$ in $\bar{\omega} \omega_0 \Delta x_{ij} = 2 \pi (\Delta x_{ij}/\lambda_{\rm gw})$ in the expansion of $f_{ij}^{\rm GW}$ given in Eq.~\eqref{expansion1}. Further, the next-to-leading order (precisely, second-order) term $\tilde{f}^{(3)}_{ij}$ sets the maximum length of the array, with interatomic spacing $\lambda_0$, within which the atoms show GW-induced cooperation (as discussed following Eq.\,\eqref{expansion1} and in Appendix\,\ref{apSec:crossover}). In the local Lorentz coordinates, the corrections to flat metric appear at $\order{(\Delta x_{ij}/\lambda_{\rm gw})^2}$ and the GW changes the separation between $i$th and $j$th atoms, to leading (zeroth) order in $\Delta x_{ij}/\lambda_{\rm gw}$, as~\cite{Maggiore2007}
\begin{equation}\label{eq:LLF-geodeviation}
    \Delta x_{ij} \to \Delta x_{ij} \left( 1 + \frac{h_+}{2} \cos\omega t \right).
\end{equation}
Any additional effects of the GW spacetime on the polarization of the EM field, e.g., rotation of the plane of polarization, are also suppressed as $\order{(\Delta x_{ij}/\lambda_{\rm gw})^2}$~\cite{Shoom2024}. Therefore, making the substitution~\eqref{eq:LLF-geodeviation} in Eq.~\eqref{eq:EM-dcr} and expanding the resulting expression to first order in $h_+$ correctly reproduces $\tilde{f}^{(1)}_{ij}$. Note that this procedure is the same as usually employed to obtain an intuitive understanding of the response of bar or interferometric detectors~\cite{Maggiore2007}. The above procedure gives
\begin{multline}
    \begin{split}
    F_{ij} &= \frac{3}{2}\Bigg\{(1 - (\vu{d}_{\rm a} \cdot\vu{x}_{ij})^2) \left[ \frac{\sin(\omega_0 \Delta x_{ij})}{\omega_0 \Delta x_{ij}} + \frac{h_+ \cos(\omega t)}{2}\left( \cos(\omega_0 \Delta x_{ij}) -\frac{\sin(\omega_0 \Delta x_{ij})}{\omega_0 \Delta x_{ij}}\right) \right] \\
    &+ (1 - 3 (\vu{d}_{\rm a} \cdot\vu{x}_{ij})^2) \Bigg[\frac{\cos(\omega_0 \Delta x_{ij})}{\omega^2_0 \Delta x^2_{ij}}-\frac{\sin(\omega_0 \Delta x_{ij})}{\omega^3_0 \Delta x^3_{ij}}
    + \frac{h_+ \cos(\omega t)}{2} \left(-\frac{\sin(\omega_0 \Delta x_{ij})}{\omega_0 \Delta x_{ij}} + 3\frac{\sin(\omega_0 \Delta x_{ij})}{\omega_0^3 \Delta x_{ij}^3} - 3\frac{\cos(\omega_0 \Delta x_{ij})}{\omega^2_0 \Delta x^2_{ij}}\right)\Bigg]\Bigg\}\\
    &+ \order{\left(\frac{\Delta x_{ij}}{\lambda_{\rm gw}}\right)^2}.
    \end{split}
\end{multline}

Specifically, the GW-induced correction to leading order in $\Delta x_{ij}/\lambda_{\rm gw}$ (the regime of GW-induced superradiance) is
\begin{multline}\label{eq:GWi-dcr-EM}
    f^{\rm GW}_{ij} = \frac{3}{2} \frac{h_+ \cos(\omega t)}{2} \Bigg\{(1 - (\vu{d}_{\rm a} \cdot\vu{x}_{ij})^2)\left( \cos(\omega_0 \Delta x_{ij}) -\frac{\sin(\omega_0 \Delta x_{ij})}{\omega_0 \Delta x_{ij}}\right) \\
    + (1 - 3 (\vu{d}_{\rm a} \cdot\vu{x}_{ij})^2) \left(-\frac{\sin(\omega_0 \Delta x_{ij})}{\omega_0 \Delta x_{ij}} + 3\frac{\sin(\omega_0 \Delta x_{ij})}{\omega_0^3 \Delta x_{ij}^3} - 3\frac{\cos(\omega_0 \Delta x_{ij})}{\omega^2_0 \Delta x^2_{ij}}\right)\Bigg\}.
\end{multline}
Once again, the total dissipative coupling obtained in the scalar-light model is recovered
for $(\vu{d}_{\rm a} \cdot\vu{x}_{ij})^2 = 1/3$:
\begin{equation}
    F_{ij} = \frac{\sin(\omega_0 \Delta x_{ij})}{\omega_0 \Delta x_{ij}} + \frac{h_+ \cos(\omega t)}{2}\left( \cos(\omega_0 \Delta x_{ij}) -\frac{\sin(\omega_0 \Delta x_{ij})}{\omega_0 \Delta x_{ij}}\right) .
\end{equation}

For general orientation of the atomic dipole moments w.r.t. the length of the array and $\omega_0 \Delta x_{ij} = 2 \pi (i-j)$, the near-field terms lead to a correction to $\eta$ given as:
\begin{equation}
    \eta_{\rm near-field} = -\frac{3}{8 \pi^2 N} \sum_{i,j;i \neq j}^{N} \frac{1}{(i-j)^2} = -\frac{1}{8},
\end{equation}
that does not scale with the number of atoms. Thus the GW-induced superradiance regime is not affected in the full EM model.
%%%%%%%%%%%%%%%%%%%%%%%%%%%%%%%%%%%%%%%%%%%%%%%%%%%%%%%%%%%%%%%%%%%%%%%%%%%%%%%%%%%%%%%%%%%%%%%%%%
\subsection{Dephasing due to coherent dipole-dipole interactions}\label{SMSec:dephasing-cdd}
\begin{figure}
    \centering
    \includegraphics[width=0.9\linewidth]{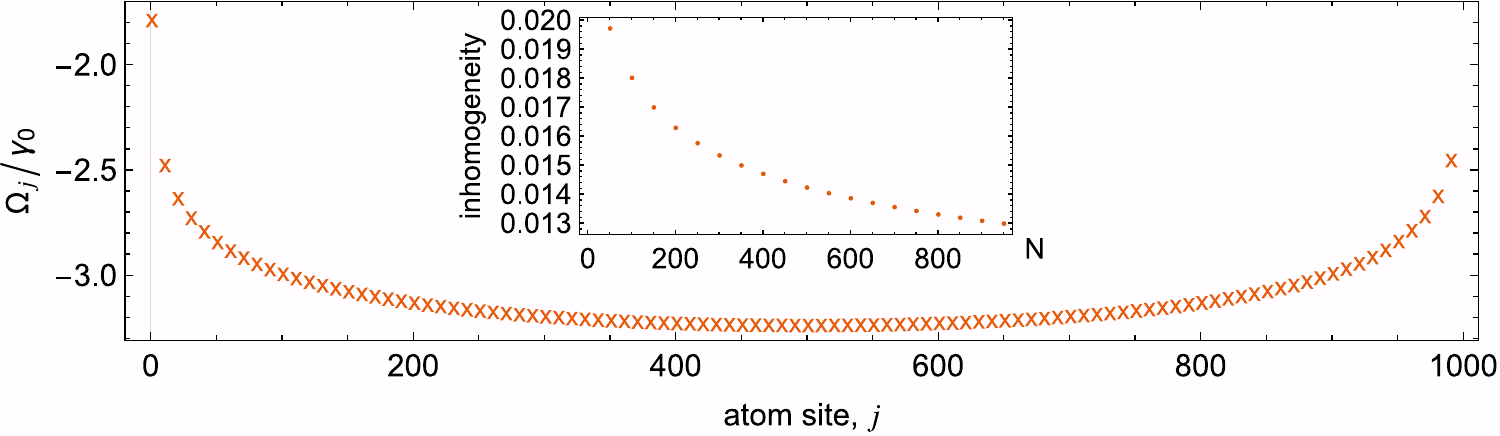}
    \caption{Collective frequency shift (normalized by the spontaneous emission rate of a single atom) at each atom site due to coherent Hamiltonian dynamics in a one-dimensional array of 1000 atoms, with $d=\lambda_0$. The inset shows that the inhomogeneity in frequency shifts, defined as the variance of frequency shifts at each site normalized by the mean frequency shift, decreases with the total number $N$ of atoms in the array. See Sec.~\ref{SMSec:dephasing-cdd} for discussion.}
    \label{fig:CFS}
\end{figure}
Similar to as done above, we can start with the expression of the flat spacetime coherent dipole-dipole coupling rate in the full EM model~\cite{Gross1982,Garcia2017,Carmichael2000}:
\begin{equation}
        \Omega_{ij} 
        = \frac{3\gamma_0}{2}\Bigg\{ -\frac{\cos (\omega_0 \Delta x_{ij})}{ \omega_0 \Delta x_{ij}} \left[1 - (\vu{d}_{\rm a} \cdot\vu{x}_{ij})^2\right] +  \left(\frac{\cos (\omega_0 \Delta x_{ij})}{ (\omega_0 \Delta x_{ij})^3} + \frac{\sin (\omega_0 \Delta x_{ij})}{(\omega_0 \Delta x_{ij})^2}\right) \left[1 - 3 (\vu{d}_{\rm a} \cdot\vu{x}_{ij})^2\right]\Bigg\},
\end{equation}
and deduce the GW-induced coherent coupling in the GW-induced superradiance regime.
The coherent Hamiltonian dynamics of the array causes Lamb shift and collective frequency shift ($\Omega_j \equiv \sum_{i\neq j} \Omega_{ij}$) for each atom~\cite{Friedberg1973}. The inhomogeneity of the collective frequency shift along the array causes dephasing of the array and can suppress collective effects~\cite{Gross1982,Masson2022}.  For $(\vu{d}_{\rm a} \cdot\vu{x}_{ij})^2 =1/3$, the Minkowskian contribution to the coherent Hamiltonian dynamics is governed by: $ \Omega_{ij}=-\gamma_0 \cos(\omega_0 \Delta x_{ij})/(\omega_0 \abs{\Delta x_{ij}}) $~\cite{Gross1982,Friedberg1973}. Therefore, in contrast to the collective emission rate, the Minkowskian contribution to the collective frequency shifts will not be suppressed for the atom distribution prescribed in Eq.~\eqref{distribution}. As a result, the dephasing time scale due to coherent dipole-dipole interactions is set by the Minkowskian contribution since the GW contribution is suppressed by the wave's minuscule amplitude. As the atoms in the bulk of an ordered array experience similar environment, the inhomogeneity (variance of frequency shifts normalized by the mean) of the collective frequency shifts along the array decreases with the total number of atoms in the array [see Fig.~\ref{fig:CFS}]. This occurs because the fraction of atoms in the bulk (relative to those on the edges) of the array increases with the total number of atoms~\cite{Masson2022}. In ordered one-dimensional arrays, the Minkowskian coherent dipole-dipole interaction thus introduces only a slow dephasing and, in fact, it has been shown that it is rather the competition between different decay channels that causes a suppression of collective effects in such systems~\cite{Masson2020,Masson2022}. The number of different decay channels available depends on the lattice constant of the array. In our analysis, this effect is contained in $\abs{\mu} N$, the effective number of cooperating atoms. Therefore, the slow dephasing of the atoms due to coherent dipole-dipole interactions is expected to give only a subleading correction to our analysis in the regime of interest. We have focused on the dissipative dynamics as it is this aspect of the dynamics that decides the fate of collective effects in ordered atom arrays~\cite{Masson2020,Masson2022}.
%%%%%%%%%%%%%%%%%%%%%%%%%%%%%%%%%%%%%%%%%%%%%%%%%%%%%%%%%%%%%%%%%%%%%%%%%%%%%%%%%%%%%%%%%%%%%%%%%%
%%%%%%%%%%%%%%%%%%%%%%%%%%%%%%%%%%%%%%%%%%%%%%%%%%%%%%%%%%%%%%%%%%%%%%%%%%%%%%%%%%%%%%
\section{Testing for vacuum-mediated GW-induced Superradiance}
As an outlook on the GW-induced collective effects mediated by the electromagnetic vacuum presented in the main text, we now explore the possibility of testing these joint effects of quantum mechanics and general relativistic gravity. Our goal here is to establish in-principle observability of these joint effects, rather than to optimize for competitive GW astronomy. We emphasize this because the two goals lead to different design choices: conventional GW sensing benefits from a strong coherent baseline EM field, whereas the simplest and cleanest setting for isolating \textit{joint} GR-quantum effects in the present context relies on the EM vacuum as the baseline, so that the predicted signal demands a joint GR and quantum-optical description. The vacuum-mediated GW-induced superradiance falls within the latter category and is our focus here.

Consider an ordered array of $N$ atoms with interatomic spacing $d=\lambda_0$.
The atomic transition linewidth would be required to satisfy $\gamma_0 \lesssim \omega$, so that the GW-induced sidebands at $\abs{\omega_{0} \pm \omega}$ can be resolved in principle.
A natural strategy for rejecting common-mode technical noise would exploit the orientation dependence of the GW-induced dissipative coupling. As discussed following Eq.~\eqref{wightman}, the angular integration of the GW contribution to the Wightman function results in $J_2(l k_\perp)$, whose sign flips between arrays oriented along $\hat{x}$ and $\hat{y}$. Consequently, two co-located arrays of identical composition---one along $\hat{x}$, the other along $\hat{y}$---yield GW-induced corrections of opposite sign, $\Delta\Gamma_\downarrow^{(x)} = -\Delta\Gamma_\downarrow^{(y)}$, under a $+$-polarized GW. In contrast, the flat-spacetime incoherent emission, laser frequency noise, blackbody radiation shifts, lattice intensity fluctuations, and Zeeman shifts are orientation-independent and therefore common to both arrays. Forming the differential observable $\mathcal{D}(t) \equiv \Gamma_\downarrow^{(x)}(t) - \Gamma_\downarrow^{(y)}(t) = 2\,\Delta\Gamma_\downarrow^{(x)}(t)$ thus doubles the signal while canceling all common-mode contributions, leaving only uncorrelated noise, predominantly photon shot noise. The complementary sum $\Gamma_\downarrow^{(x)}(t) + \Gamma_\downarrow^{(y)}(t) - 2\,\Gamma_\downarrow^{\mathrm{inc}}(t)$ provides a null channel that vanishes for a pure $+$-polarized wave and can serve as a diagnostic of residual systematics. This orientation-dependent response is ultimately a consequence of the spin-2 character of the gravitational wave and is therefore intrinsic to the GW-matter coupling rather than an artifact of a particular detection scheme. Further, unlike the GW-induced emission, which is modulated at the GW frequency, the flat-spacetime contribution to the photon emission intensity remains unmodulated. Therefore, the latter contribution can in principle be filtered out through lock-in detection or matched filtering~\cite{Allen2012,Maggiore2007}, leaving only the shot-noise contribution from it.

Therefore, from the perspective of the mechanism underlying GW-induced superradiance, the key noise sources (for ascertaining in-principle observability) can be identified as (i) positional disorder in the atom array, which could suppress the GW-induced superradiance or re-enable the flat-spacetime superradiance that we intentionally suppress, (ii) atoms missing from lattice sites, and (iii) photon shot noise in the readout of GW-induced sidebands. As shown in Fig.~\ref{fig:mungw-noisy}, the GW-induced superradiance is robust against position disorder and partial filling in atom arrays. Left with the noise source (iii), we explicitly take it into account to compute the signal-to-noise ratio for testing vacuum-mediated GW-induced superradiance. To this end, consider the Fourier components of the photon emission rate: $\mathcal{S}(\nu) = \frac{1}{2 \pi} \int_{- \infty}^{\infty} \dd{t} e^{i (\nu - \omega_0)} \Theta(t) \Gamma_{\downarrow}(t)$,
where $\Theta(t)$ is the Heaviside theta function, enforcing the fact that photon emission follows the exciting pulse at $t=0$.
The Fourier transforms of the flat-spacetime and the dominant GW-induced contributions yield, in the $\omega \gg \gamma_0$ limit: $\mathcal{S}_{\rm M}(\nu=\omega^+_0) = (N/2\pi) (\gamma_0/\omega)^2$ and $\Delta \mathcal{S} (\nu=\omega^+_0) \approx \mu_{\rm GW}(0)N^2/8\pi$, respectively. Corresponding to the differential observable discussed above, the signal-to-noise ratio is:
\begin{equation}
	\mathrm{SNR}\approx \frac{\abs{\mu_{\rm GW}(0)}N^2}{4 \sqrt{\pi}} \frac{\omega/\gamma_0}{\sqrt{N}}.
\end{equation}
As long as the total number of atoms in an array with $d=\lambda_0$ is less than the threshold for GW-induced coupling identified in the main text, we have $\mu_{\rm GW}(0)N^2 \approx h_+ N^2/2$, giving $\mathrm{SNR} \approx h_+N^{3/2} \omega/(8\sqrt{\pi}\gamma_0)$.
Equivalently, for a target SNR of unity, we can obtain the minimum detectable strain:
\begin{equation}
    h^{\rm min}_+ \approx \frac{8 \sqrt{\pi}}{N^{3/2}} \frac{\gamma_0}{\omega}.
\end{equation}
Recall that the GW frequency to which an array with $N$ atoms has highest sensitivity is $\bar{\omega}_{\rm max} \approx 1/2N$. For example, if $N=10^6$ then $f_{\rm gw, max} = \omega_{\rm max}/2\pi \approx 215$MHz, and therefore at highest sensitivity of such an array we get $h^{\rm min}_+ \approx 6.6 \times 10^{-20}$. On the other hand, if we fix $f_{\rm gw}=\omega/2\pi =100$MHz, $N=10^7$, we get $h^{\rm min}_{+} \sim 4.5 \times 10^{-21}$. For these examples, we have considered an atomic transition linewidth $\gamma_0/2\pi \sim 10^{-3}\,$Hz, corresponding to the $\leftindex^3 {P}_0-\leftindex^1 {S}_0$ clock transition at 698\,nm in $\leftindex^{87}\,{\rm Sr}$~\cite{Norcia2016,Norcia2018prx}.
The preceding discussion serves to show that fundamental noise does not trivially nullify the GW-induced collective signal. The scaling of the SNR points to the ultra-high frequency coherent monochromatic GWs in the MHz-GHz frequency range~\cite{NAggarwal2025} as the most favorable for testing the vacuum-mediated free space GW-induced superradiance with $N\sim 10^6-10^7$. Searches for MHz-GHz GWs are an active frontier for probing primordial and beyond-Standard-Model physics beyond conventional astrophysical sources~\cite{NAggarwal2025,Berlin2022,Reina2025,Fischer2025}. Note that the above-mentioned atom number requirement is for testing the free space vacuum-mediated effect and should be viewed in the context of testing a joint effect of GR and many-body quantum optics. The GW-induced dissipative coupling can be boosted inside a cavity or by using non-vacuous states of the EM field which could lower the requirement on the number of atoms. Moreover, the requirement is a total number of coherently participating emitters, not necessarily a single one-dimensional chain; higher-dimensional ordered geometries could reduce the physical extent of the setup, although their detailed collective mode structure requires separate analysis. 
   
The requirement for large coherent atom arrays for testing the vacuum-mediated GW-induced superradiance overlaps with the needs of quantum information processing and optical lattice clock applications. The design and development of systems of both the kinds have advanced rapidly over the past decade. Ordered arrays containing up to $6 \times 10^3$ atoms have already been demonstrated~\cite{Norcia2024,Manetsch2025}, with arrays of $10^4$ atoms well within reach~\cite{Manetsch2025}. Further, a 1D (cavity-based) optical lattice clock employing approximately $10^5$ atoms (not as an ordered array) with tens of seconds of atomic coherence has already been put to use to sense gravitational redshift~\cite{Bothwell2022}, and coherence time exceeding $100$\,s on the Sr clock transition has been recently demonstrated~\cite{Shuo2025,Kyungtae2025}.
Interestingly, there are several promising directions for further investigation into strategies for relaxing the required number of atoms. One possibility is to study the sensitivity of flat-spacetime subradiant states~\cite{Garcia2017} to GW-induced modification of the electromagnetic Green's function. Such states can boost the signal-to-noise ratio by strongly suppressing the flat-spacetime incoherent photon emission from the atoms, potentially lowering the number of atoms required.
Another promising avenue to improve the signal-to-noise ratio is provided by a cavity. Though we have considered free space superradiance, generalization of the results to cavity-QED systems is plausible with appropriate modifications. One important distinction compared to the free space case is that a cavity already enables all-to-all dissipative coupling in flat spacetime. Consequently, the flat-spacetime and the GW-induced collective photon emission cannot be distinguished through a choice of interatomic spacing.
Regardless, the two superradiant regimes can still be demarcated inside a cavity by tuning the cavity frequency $\omega_{\text{c}}$ to either the carrier or sideband frequencies. The GW-induced superradiance regime would be cleanly isolated for $\omega_{\rm c} \approx |\omega_0 \pm \omega|$ and $\omega \gg \omega_{\rm c}/Q_{\rm c}$, where $Q_{\rm c}$ is the cavity's quality factor.
%%%%%%%%%%%%%%%%%%%%%%%%%%%%%%%%%%%%%%%%%%%%%%%%%

%%%%%%%%%%%%%%%%%%%%%%%%%%%%%%%%%%%%%%%%%%%%%%%%%%%%%%%%%%%%%%%%%%%%%%%%%%%%%%%%%
        
\end{document}